\documentclass{aa}
\usepackage{times}
\usepackage{natbib}
\usepackage[dvips]{graphicx}
\usepackage{psfig,longtable,lscape}
\usepackage{rotating}
\bibpunct{(}{)}{;}{a}{}{,}

\def\be{\begin{equation}}
\def\ee{\end{equation}}
\def\bea{\begin{eqnarray}}
\def\eea{\end{eqnarray}}
\begin{document}

\def\ltsima{$\; \buildrel < \over \sim \;$}
\def\simlt{\lower.5ex\hbox{\ltsima}}
\def\gtsima{$\; \buildrel > \over \sim \;$}
\def\simgt{\lower.5ex\hbox{\gtsima}}

\title{XMM--Newton and ESO observations of the two Unidentified $\gamma$--ray Sources 3EG J0616$-$3310 and
3EG J1249$-$8330\thanks{This work is based on observations obtained with {\em XMM-Newton}, an ESA
science mission with instruments and contributions directly funded by ESA Member States and NASA. Optical observations have been obtained with the ESO/MPG 2.2m at La Silla (Chile) under programs  68.D--0478(A) and  68.D--0478(B).}}

\author{N. La Palombara\inst{1}, R. P. Mignani\inst{2}, E. Hatziminaoglou\inst{3}, M. Schirmer\inst{4}, G.F. Bignami\inst{5,6}, P. Caraveo\inst{1}}

\institute{INAF -- IASF Milano, Via E. Bassini 15, I--20133 Milano (I)
         \and
         	Mullard Space Science Laboratory, University College London, Holmbury St Mary,
Dorking, Surrey RH5 6NT (UK)
         \and
	Instituto de Astrofisica de Canarias, Via Lactea, E--38200 La Laguna--Tenerife (E)
         \and
         	Isaac Newton Group of Telescopes, Edificio Mayantigo, Calle Alvarez Abreu 68, E--38700 Santa Cruz de la Palma (E)
         \and
         	Centre d'\'Etude Spatiale des Rayonnements (CESR), CNRS--UPS, 9 Avenue du colonel Roche, F-31028 Toulouse (F)
	\and
	Universit\`a di Pavia, Dipartimento di Fisica Teorica e Nucleare, Via Ugo Bassi 6, I--27100 Pavia (I)
	}

\authorrunning{N. La Palombara et al.}

\titlerunning{X--Ray And Optical Coverage Of Two UnID $\gamma$--Ray Sources}

\abstract{The limited  angular resolution of  $\gamma$--ray telescopes
prevents a  direct identification of the majority  of sources detected
so  far.  This  is particularly  true for  the low  latitude, probably
galactic,  ones  only  10  \%  of which  has  been  identified.   Most
counterparts of the identified low--latitude $\gamma$--ray sources are
{\it Isolated  Neutron Stars} (INS),  both {\em radio--loud}  and {\em
radio--quiet} (Geminga--like)  objects, which are  characterised by an
extremely  high value of  the \textit{X--ray--to--optical}  flux ratio
$f_{\rm X}/f_{\rm opt}$.  Therefore, the systematic X--ray and optical
coverage of low--latitude  unidentified $\gamma$--ray sources aiming at
high $f_{\rm X}/f_{\rm  opt}$ sources seems one of  the most promising
ways to spot INS candidate  counterparts.  Since low  latitude sources
are heavily affected by the  interstellar absorption at both X--ray and
optical  wavelengths,  we   have  focussed  on  two  middle--latitude,
probably galactic,  {\em GRO/EGRET} sources: 3EG  J0616$-$3310 and 3EG
J1249$-$8330.   These  two sources,  which  could  belong  to a  local
galactic population, have been selected owing to their relatively good
positional   accuracy,   spectral   shape   and  lack   of   candidate
extragalactic   radio   counterparts.   Here   we   report  on   X-ray
observations of the two  $\gamma$--ray error boxes performed with {\em
XMM--Newton} and on their optical  follow--up carried on with the {\em
Wide Field Imager}  at the ESO/MPG 2.2m telescope.   Less than half of
the $\sim$ 300 sources detected by the X--ray coverage have no optical
counterparts.  Among  those, we have selected  few interesting sources
with $f_{\rm  X}/f_{\rm opt}\ge 100$, which we  consider promising INS
candidates.   }

\maketitle

\section{Introduction}\label{sec:1}

The nature of  the Unidentified $\gamma$--ray Objects (UGOs)  is one of
the main issues of the $\gamma$--ray astronomy.  The third {\em Compton
Gamma   Ray   Observatory}   ({\em   CGRO})  {\em   EGRET}   catalogue
\citep{Hartman+99}  contains   271  high--energy  $\gamma$--ray  sources
detected  at   energies  above   0.1  GeV.   The   high--latitude  ones
($|b|>10^{\circ}$) are 191,  and 93 of them have  been identified with
{\em blazars}  i.e.  featureless flat  spectrum radio--loud AGN  and BL
Lac objects \citep{vonMontigny+95},  while 5 of the 80  sources at low
latitude  have been  identified with  pulsating {\em  Isolated Neutron
Stars}   (INSs),  both   classical   radio--pulsars  and   radio--quiet,
Geminga--like,  objects \citep{Caraveo02,Thompson04a}.   In  total, 170
{\em EGRET} sources,  74 of which at low  latitudes, have no
counterpart  at  lower  frequencies  and remained  unidentified.   The
identification work had been  hampered mainly by the poor localization
(about   1$^{\circ}$  in   diameter   at  low--latitudes   and  up   to
1.5$^{\circ}$  at  mid--latitudes)  which  frustrated  the  search  for
counterparts  at other wavelengths. Moreover, the limited $\gamma$--ray
statistics made it impossible  to perform `blind' periodicity analysis
aimed at unveiling undiscovered INSs.

\begin{figure}[t]
\begin{center}
\resizebox{\hsize}{!}{\includegraphics[angle=+90,clip=true]{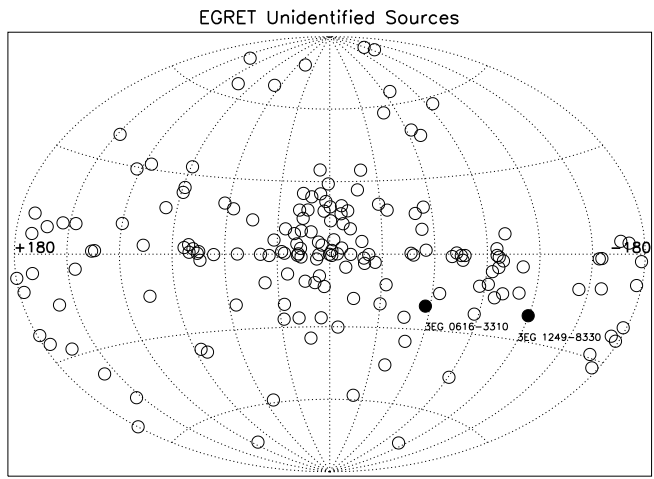}}
\end{center}
\caption{Galactic distribution  of  the  unidentified $\gamma$--ray sources from the third {\em EGRET} catalogue (Hartman et al. 1999):  the two black circles show the  positions of 3EG J0616$-$3310 and 3EG J1249$-$8330}\label{galacticposition}
\end{figure}
\vspace{-0.25 cm}

The   distribution  analysis of  the  UGOs in  the third  {\em
EGRET} catalogue (Figure~\ref{galacticposition}) shows that they can be grouped in at least
four                         different                        populations
\citep{Gehrels+00,Grenier00,Grenier01,Grenier04,Romero+04}.
The presumably galactic UGOs (included those observed
at $|b|>10^{\circ}$) are  equally  distributed  in 3
populations, with about 45 sources each. The  GRP--I includes bright and  relatively hard sources
near  the  galactic  plane  ($|b|<3^{\circ}$),  with  a  concentration
towards the inner spiral  arms \citep{Bhattacharya+03} at distances of
a  few kpc  \citep{Kanbach+96,Romero+99}.  Many  of these  sources are
well correlated with tracers of star formation, which means that their
age should be a  few million years at most \citep{Romero01,Grenier04}.
The  GRP--II includes  variable sources  at $|b|>3^{\circ}$  which are
distributed in  a sort  of spherical halo  around the  Galactic Center
with  a radial distribution  equivalent to  that of  globular clusters
\citep{Grenier01,Grenier04}.     These   sources   are    softer   and
significantly      more     variable      than      GRP--I     sources
\citep{Torres+01a,Torres+01b,Nolan+03};  they   are  presumably  older
(with an age  of the order of  a few Gyrs) and at  distances between 2
and  8  kpc.   The  last   group  is the {\em  Local  Gamma--Ray
Population}    (LGRP). It  is    composed   by    stable    sources   at
$3^{\circ}<|b|<30^{\circ}$, which are  fainter and softer than sources
located  at lower  Galactic latitude  and have  an  evident asymmetric
distribution north of the Galactic center and south of the anticenter.
\citet{Gehrels+00} and \citet{Grenier00}  suggested that these sources
might be associated with the Gould  Belt, i.e.  a 30 Myr old starburst
region,  300 pc in  radius, composed  by young  massive and  late type
stars,   molecular   clouds,   and  expanding   interstellar   medium.
Therefore, LGRP sources  should be young (i.e.  a  few Myr old) nearby
(100--400  pc) and  low  luminosity ($\sim  10^{32-33}$ erg  s$^{-1}$)
objects \citep{Grenier04}. But recently, using an improved interstellar emission model, \citet{Casandjian+05} have shown that most of these sources can correspond to the emission of clumpy dark clouds that surround all the molecolar clouds of the Gould Belt; therefore their existence is not confirmed. Finally, there is an isotropic population  of extragalactic  origin, which is characterized by a variety  of spectra and variabilities and includes  no more than 35 sources.

After  the release  of the  third {\em  EGRET} catalogue,  only  a few
 additional   identifications   have    been   obtained.    From   the
 extragalactic point of view,  the multiwavelength approach has led to
 a    blazar    identification of   3EG    J2016+3657
 \citep{Halpern+01c},   3EG  J2006-2321  \citep{Wallace+02}   and  3EG
 J2027+3429    \citep{Sguera+04},     while    for    3EG    1621+8203
 \citep{Mukherjee+02},   3EG  J1735-1500   \citep{Combi+03}   and  3EG
 J0416+3650  \citep{Sguera+05} a  radio--galaxy has  been  proposed as
 counterpart.  Within the  Galaxy, 3EG  J1824-1514 and  3EG J0241+6103
 have been  associated with the  two well--known microquasars  LS 5039
 \citep{McSwain+04} and LS I +61 303 \citep{Massi04,Casares+05}, while
 \citet{Combi+04} suggested the association between the microquasar
 candidate AX J1639-4642 and the UGO 3EG J1639-4702.  In two cases a
 peculiar early--type binary system have been suggested as counterpart
 of an UGO, i.e.   SAX J0635+0533 for 3EG J0634+0521 \citep{Kaaret+00}
 and  A0535+26  for 3EG  J0542+2610  \citep{Romero+01}. 
 Turning now to the INS family, the search for radio or X--ray pulsars
 has provided likely candidates for
 3EG J0222+4253 (PSR J0218+4232, \citet{Kuiper+02}), 3EG J1048-5840
 (PSR B1046-58, \citet{Kaspi+00}), 3EG J2021+3716
 (PSR J2021+3651, \citet{Roberts+02})  and  3EG  J2227+6122
 (RX/AX J2229.0+6114, \citet{Halpern+01a,Halpern+01b}).   Moreover,  thanks  to  the  Parkes
 Multi--beam   pulsar    Survey   (PMS)   of    the   galactic   plane
 \citep{Manchester+01},  other promising  pulsar candidates  have been
 found for  3EG J1420-6038  and 3EG J1837-0606  \citep{D'Amico+01} and
 for  3EG J1013-5915  \citep{Camilo+01}.
 However, none of the proposed identifications could be confirmed owing
 to the lack of contemporary radio data. Finally,  various  X--ray and
 radio studies have pointed to a  close relationship between
 the  {\em pulsar  wind nebulae}  (PWN)  and the  {\em EGRET}  sources
 \citep{RobertsGaenslerRomani02,Roberts+01a,Roberts+04b,Roberts05}.

Apart from the cases of Cen A and of the LMC, all the {\em  EGRET} sources identified so far fall either in the INS
or in the  Blazar class. However, also in view of the \textit{High Energy
Stereoscopic  System}  (HESS)  results  which  are  unveiling  several
classes     of    very     high    energy     $\gamma$--ray    emitters
\citep{Aharonian+05a,Aharonian+05b,Aharonian+05c,Aharonian+05d,Brogan+05},   it   is
hard  to   believe  that  these  two  classes   exhaust  all  possible
$\gamma$--ray source  scenarios.  Nevetheless, it has  been argued that
rotation  powered pulsars  should dominate  the  Galactic $\gamma$-ray
source population  \citep{YadigarogluRomani97} and that  many of those
should be radio-quiet, since the $\gamma$-ray beam is broader than the
radio one.   The classic  example of a  radio--quiet pulsar  is Geminga
\citep{Caraveo+96},  which  offers   an  elusive  template  behaviour:
prominent in  high energy  $\gamma$--rays, easily detectable  in X--rays
but downright  faint in  optical, with sporadic  or no  radio emission
\citep{BignamiCaraveo96}.  In the latest years, the idea that galactic
UGOs  might   be  associated   with  Geminga--like  objects   has  been
strengthened  by  the  cases  of  three UGOs  which  were  associated with radio--quiet INSs:  3EG J1835+5918, the brightest among
the   unidentified   UGOs    and   considered   `the   next   Geminga'
\citep{MirabalHalpern01,Reimer+01,Halpern+02}, and the two sources 3EG
J0010+7309  and  3EG  J2020+4017,  positionally  coincident  with  the
supernova    remnants    CTA--1   \citep{Brazier+98,Halpern+04}    and
$\gamma$--Cygni \citep{Brazier+96}, respectively.

\begin{table*}[!ht]
\caption{Main characteristics of the eight {\em EPIC} observations of 3EG J0616$-$3310 (ID = 1--4) and 3EG  J1249$-$8330 (ID = 5--8)}\label{observations}
\begin{center}
\footnotesize{
\begin{tabular}{|c|ccccccccc|} \hline
Field & Rev. & Date & 	R.A. (J2000)	&	DEC (J2000) & \multicolumn{3}{c}{Exposure Time (ks)} & $N_{\rm H}$ & Detected \\
ID   &      &    (UT)   & $^h$ $^m$ $^s$ 	& $^\circ$ ' ''	 & PN & MOS1 & MOS2 & $(10^{20}$ cm$^{-2})$ & Sources \\ \hline
1 & 346 & 2001-10-29T17:04:09 & $06$ $17$ $47.1$ & $-32$ $55$ $13.9$ & 6.8 & 11.4 & 11.5 & 2.7 & 50 \\
2 & 341 & 2001-10-18T23:53:02 & $06$ $17$ $47.1$ & $-33$ $25$ $13.9$ & 6.7 & 12.0 & 12.0 & 2.5 & 37 \\
3 & 346 & 2001-10-29T04:27:17 & $06$ $15$ $24.1$ & $-33$ $25$ $13.9$ & 2.5 & 7.3 & 7.7 & 2.4 & 32 \\
4 & 346 & 2001-10-28T23:26:57 & $06$ $15$ $24.1$ & $-32$ $55$ $13.9$ & 1.3 & 6.3 & 7.7 & 2.5 & 27 \\
5 & 236 & 2001-03-23T12:56:43 & $12$ $57$ $53.1$ & $-83$ $15$ $01.9$ & 7.0 & 11.2 & 11.3 & 10.2 & 38 \\
6 & 236 & 2001-03-23T17:54:20 & $12$ $57$ $53.1$ & $-83$ $45$ $01.9$ & 8.2 & 11.2 & 10.9 & 8.4 & 51 \\
7 & 239 & 2001-03-30T03:50:11 & $12$ $40$ $13.1$ & $-83$ $45$ $01.9$ & 0.8 & 2.9 & 2.4 & 9.4 & 7 \\
8 & 239 & 2001-03-29T22:28:14 & $12$ $40$ $13.1$ & $-83$ $15$ $01.9$ & 8.3 & 12.7 & 12.9 & 11.2 & 52 \\ \hline
\end{tabular}}
\end{center}
\end{table*}
\vspace{-0.25 cm}

The energetic of the Geminga--like objects is not sufficient to account
for the very low latitude, presumably more distant, GRP--I sources but
could account  for several middle latitude, rather  nearby, objects of
the LGRP.  Based on this rationale, we considered the case of two
middle  latitude (see Figure~\ref{galacticposition}) UGOs:
3EG  J0616-3310  and   3EG  J1249-8330  (\citet{LaPalombara+05}, Paper I).   Both
sources are  characterized by  a relatively good  positional accuracy,
with      a     95\%      confidence      error     circle      radius
$\theta_{95}\sim0.6^{\circ}$,   by   a   power   law   photon   index
$\Gamma\sim2.1$ and by a steady  emission with fluxes above 0.1 GeV of
(12.6 $\pm$ 3.2)   and    (19.9 $\pm$ 4.4) $\times10^{-8}$   photon   cm$^{-2}$
s$^{-1}$, respectively \citep{Hartman+99}. In both  cases, the lack of radio counterparts
down     to     a    limit     of     a     few     tens    of     mJy
\citep{Mattox+01,Tornikoski+02,Sowards--Emmerd+04} does  not support an
extragalactic identification, while an association with the Gould Belt
appears more likely (their existence was confirmed also by \citet{Casandjian+05}). Thus we have applied the multiwavelength approach
successfully used  for the  identification of Geminga  as well  as for
other       UGOs      associated       with       radio--quiet      INS
\citep{Caraveo2001,MukherjeeHalpern04}.    In   the   multiwavelength
approach,  we  start  with  the  X--rays since  they  are  the  nearest
neighbours to $\gamma$--rays and can  be used to bridge the gap between
the poor  resolution achievable in $\gamma$--rays and  the standards of
optical or radio astronomy. First, an X--ray image of the
$\gamma$--ray  error--box is taken  and the  detected X--ray  sources are
cross--correlated with  optical and radio  catalogues, either available
in archives  or produced from \textit{ad hoc} observations.  Next, potential  INS candidates are singled out  amongst X--ray sources
with  an extremely  high  value of  the  \textit{X--ray--to--optical} flux  ratio
$f_{\rm X}/f_{\rm opt}$.  This `top--down' systematic strategy is
a time  consuming exercise since it  usually  involves several
observing cycles  with different instruments  at different facilities,
both space and ground--based. Thus it can be pursued for a
large  number  of  $\gamma$--ray sources  only  if  a  semi--automatic
procedure is set--up.  Our work can be viewed as a preparatory step for
the massive  identification work  which will be  needed in  the coming
years,   when   {\em   AGILE}   \citep{Tavani+03}  and   {\em   GLAST}
\citep{McEnery+04} will deliver  hundreds of new $\gamma$--ray sources.
The improved angular resolution of these new $\gamma$--ray telescopes will
provide arcmin  positioning, thus easing  considerably the identification
work.   To   this   aim,    a   statistical approach   will   be   necessary
\citep{TorresReimer05}.

X--ray observations  and data  reduction techniques are  presented in
\S\ref{sec:2},  while the  X--ray source analysis  is described  in
\S\ref{sec:3}.   The   optical  observations  are described  in \S\ref{sec:4}, while the
cross--correlations of the X--ray data with the optical and radio ones are described in \S\ref{searchWFI}. In \S\ref{sec:6} we discuss the source X--ray/optical analysis, while the  most interesting  sources are
presented in \S\ref{sec:7}. Finally, in \S\ref{sec:8} summary and conclusions are outlined.
\vspace{-0.25 cm}

\section{X--ray observations}\label{sec:2}

%\subsection{X-ray Observations}

The error boxes  of 3EG J0616$-$3310 and 3EG  J1249$-$8330 are circles
of $\sim$ 35 arcmin radius, a value comparable  to the field of  view of the
{\em XMM-Newton}  telescopes \citep{Jansen+01}.  Furthermore,  with an
unprecedented collecting  area of  $\sim$ 2500 cm$^2$ @  1 keV,  a good
spectral resolution  ($\sim$ 6 \% @  1 keV) and a  rather broad energy
range (0.1--10  keV), the {\em  European Photon Imaging  Camera} ({\em
EPIC}),   is  particularly  suited   to  investigate   faint  sources.
Therefore, we covered each {\em EGRET} error box with four $\simeq$ 10
ks {\em  EPIC} observations (see Paper I), corresponding to $\sim$ 70 \% of the 95 \% error circle. They did not cover the central part of the error circle, which is not a favoured region for the true position of the $\gamma$--ray source. Therefore, the probability that it was actually covered by our observations depends only on their geometry and is $\sim$ 70 \%.
In each observation  all the  three {\em
EPIC}  focal   plane  cameras  were   active:  the  {\em   PN}  camera
\citep{Strueder+01} was  operated in  Extended Full Frame  mode, while
the {\em MOS1} and  {\em MOS2} cameras \citep{Turner+01} were operated
in standard  Full Frame mode. In  all cases the thin  filter was used.
The   details    of   the    eight   observations   are    listed   in
Table~\ref{observations}  where,  for  each  of them,  we  report  the
pointing  coordinates, the  `effective'  exposure times  of the  three
cameras  after  the  soft-proton rejection  (\S\ref{processing}),  the
galactic neutral hydrogen column density in the pointing direction, and
the total number of detected sources.
\vspace{-0.25 cm}

\subsection{Data Processing}\label{processing}

For each pointing  we obtained three data sets, one  for each camera,
which  were   independently  processed  through   the  standard  {\em
XMM-Newton Science Analysis System} (SAS) v.5.2.  In the first step,
the {\em XMM/SAS}  tasks {\em emproc} and {\em  epproc} were used to
linearize the {\em  MOS} and {\em PN} event  files, respectively.  In
the second step, event files  were cleaned up for the effects of hot
pixels and soft proton flares.

Hot pixels, flickering pixels  and bad columns, which produce spurious
source detections during an observation, are largely removed by the on
board data processing software while  the remaining ones are masked by
the {\em XMM/SAS}. However, we found that several spurious events were
still present in  the processed data. These have  been localized using
the IRAF task {\em cosmicrays} and removed using the multipurpose {\em
XMM/SAS}  task {\em evselect}.   We then  filtered out  time intervals
affected by  high instrument background induced by  soft proton flares
(energies less than  a few hundred keV) hitting  the detector surface.
To this  aim, we accumulated the  light curves of  the events selected
close to the CCDs edges and with energies greater than 10 keV to avoid
contributions from  real X-ray source  variability.  For the  {\em PN}
data we considered only single pixels events (PATTERN = 0) with energies
10--13 keV while for the {\em  MOS} data we considered both single and
double  pixels events  (PATTERN $\le$ 4) with  energies 10--12.4  keV in
CCDs  2--7.   Then,  we set  a  count  rate  threshold for  good  time
intervals at 0.15 and 0.35 cts s$^{-1}$ for the {\em MOS} and the {\em
PN} data,  respectively.  By selecting only events  within the
`good' time intervals we finally  obtained three `clean' event lists for
each  observation, whose  `effective' exposure  times are  reported in
Table~\ref{observations}.  As  can be seen, the space  weather was not
favourable  during the  observation  of  field 7  (which  was so  much
affected by the  high particle background that we lost up  to 80 \% of
the integration  time) and, partly,  for the observations of  fields 3
and 4.  An  example of the processed {\em EPIC/MOS}  image is shown in
Figure 2, in the case of field 4.
\vspace{-0.25 cm}

\subsection{Source Detection}

In order  to increase the  {\em signal to  noise} (S/N) ratio  for the
detected sources and  to reach fainter X-ray fluxes,  for each observation  we used the {\em
XMM/SAS} task {\em merge} to merge the clean linearized event lists of
the 3 cameras.  This was  possible thanks to the
excellent relative astrometry  between the {\em MOS} and  the {\em PN}
($\sim$ 1$''$, a  value much smaller than the  FWHM of the PSF).   In such a
way  we obtained  a `total'  event list,  which was  used  to generate
images in 7 different energy ranges.  Namely, we considered two standard coarse
soft/hard bands  (0.5--2 and  2--10 keV) and  a finer  energy division
(0.3--0.5, 0.5--1, 1--2, 2--4.5,  4.5--10 keV).  All images were built
with a  spatial binning of  4\farcs35 per pixel, roughly  matching the
physical binning  of the  {\em PN} images  (4$''$/pixel) and  a factor
about   four  larger   than  the   one   of  the   {\em  MOS}   images
(1\farcs1/pixel).   For  each  energy  band, a  corresponding  set  of
exposure  maps  (i.e.   one  for  each of  the  three  detectors)  was
generated with  the {\em  XMM/SAS} task {\em  expmap}, to  account for
spatial  quantum  efficiency  (QE),  mirror vignetting  and  field  of
view. Finally,  the individual detector  maps were merged in  order to
obtain the  `total' exposure map  for the relevant energy  range. Both
images  and exposure  maps were  used as  a reference  for  the source
detection, which was performed in three steps:
\vspace{-0.25 cm}
\begin{enumerate}
\item For each  of the selected energy bands,  the {\em XMM/SAS} task
{\em eboxdetect} was  run in {\em local mode}  to create a preliminary
source list.   Sources were identified  by applying the  standard {\em
minimum detection likelihood}  criterium, i.e., only candidate sources
with detection  likelihood {\em -ln~P} $\ge$  5, where {\em  P} is the
probability  of  a  spurious  detection  due to  a  Poissonian  random
fluctuation of the background, were validated.
\item  Then, the  task  {\em esplinemap}  was  run to  remove all  the
validated sources from  the original image and to  create a background
map by fitting the so called {\em cheesed image} with
a cubic spline.
\item Finally,  for each  of the selected  energy bands the  task {\em
eboxdetect} was run again, but in  {\em map mode} using as a reference
the  calculated  background maps.   The  likelihood  values from  each
individual  energy  band  were  then  added and  transformed  to  {\em
equivalent single band} detection  likelihoods, with a threshold value
of 10 applied to accept or reject a detected source.
\end{enumerate}

At  the  end of  the  detection  procedure,  for each  observation  we
obtained a master source list, which provides the source counts, count
rates and detection  likelihoods for all sources in  all the detection
bands, together  with their image  position, sky position  and overall
detection likelihood.

Unfortunately, even using the maximum number of spline nodes (20), the
fit performed  in step 2 (see  above) is not  sufficiently flexible to
model  the local  variations  of the  background.   Therefore, it  was
necessary  to  correct  each  background  map {\em  pixel  by  pixel},
measuring  the counts  both  in the  {\em  cheesed image}  and in  the
background map  itself applying the correction  algorithm described in
\citet{Baldi+02}.  All sources in the master list were checked against
the  corrected background  maps  and all  their parameters  calculated
again.  Then, for all the  selected energy bands, we applied a minimum
threshold  of  8.5 on  the  corrected  detection likelihoods  $-ln~P$,
corresponding  to a probability  $P(j)=2\cdot10^{-4}$ that  the source
count  number in  the  energy  band $j$  originate  from a  background
fluctuation. This implies a contamination of at most 1 spurious source
per energy band.  The revised  source master list was then filtered to
include only sources with $-ln~P >8.5$ in at least one of the selected
energy  bands   and  manually   screened  to  reject   residual  false
detections.  The  final master  lists contain a  total of 146  and 148
sources for the  3EG J0616$-$3310 and in the  3EG J1249$-$8330 error box,
respectively.

\begin{figure}
%\vspace{8.8cm}
\begin{center}
\resizebox{\hsize}{!}{\includegraphics[angle=0,clip=true]{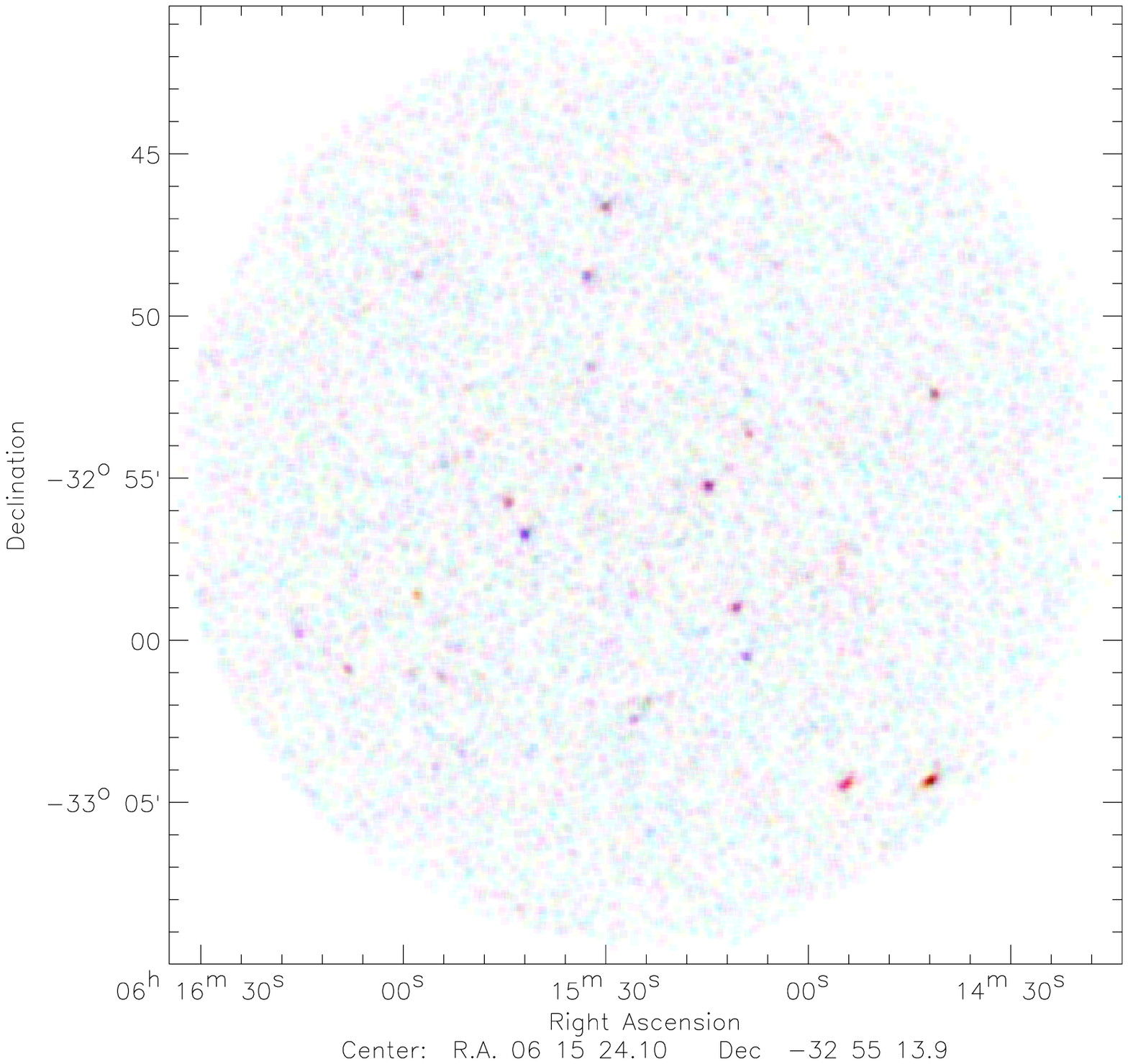}}
\end{center}
\caption{Processed XMM-Newton {\em EPIC/MOS} image of field 4 (see Table~\ref{observations}. Photons selected from three different energy bands are colour-coded: red (0.3--1 Kev); green (1--3 KeV); blue (3--10 Kev).}\label{X-ray-ima}
\vspace{-0.25 cm}
\end{figure}
\vspace{-0.25 cm}

\section{X--ray Analysis}\label{sec:3}

\subsection{Global Statistics}

In  order to  perform a  full  statistical analysis,  we computed  the
number of detected sources
in each of the 7 energy ranges, i.e.  those ones with $P(j)<2\cdot10^{-4}$ in the energy
band $j$.  Of course, most of the sources are detected in several bands:  in
Table~\ref{detections} we report, for each observation of both {\em EGRET} fields, the
number of detected sources in each energy range and its percentage over the total (that is
the total number of the sources which have been detected in at least one energy band).

\begin{table*}
\caption{X--ray sources detected in each energy range.}\label{detections}
%\vspace*{0.5truecm}
\begin{center}
\footnotesize{
\begin{tabular}{|c|cc|ccccc|c|} \hline
Range (keV) & 0.5--2 & 2--10 & 0.3--0.5 & 0.5--1 & 1--2 & 2--4.5 & 4.5--10 & Total \\ \hline
Field ID & N(\%) & N(\%) & N(\%) & N(\%) & N(\%) & N(\%) & N(\%) & N \\ \hline
1 & 37 (74) & 18 (36) & 11 (22) & 25 (50) & 29 (58) & 18 (36) & 1 (2) & 50 \\
2 & 28 (75.7) & 8 (21.6) & 6 (16.2) & 23 (62.2) & 17 (45.9) & 11 (29.7) & 0 (0) & 37 \\
3 & 28 (87.5) & 6 (18.7) & 7 (21.9) & 14 (43.7) & 21 (65.6) & 7 (21.9) & 2 (6.2) & 32 \\
4 & 26 (96.3) & 9 (33.3) & 4 (14.8) & 11 (40.7) & 14 (51.9) & 11 (40.7) & 1 (3.7) & 27 \\ \hline
3EG J0616$-$3310 & 119 (81.5) & 41 (28.1) & 28 (19.2) & 73 (50) & 81 (55.5) & 47 (32.2) & 4 (2.7) & 146 \\ \hline
5 & 32 (84.2) & 8 (21.1) & 4 (10.5) & 10 (26.3) & 20 (52.6) & 6 (15.8) & 1 (2.6) & 38 \\
6 & 37 (72.5) & 13 (25.5) & 6 (11.8) & 16 (31.4) & 29 (56.9) & 13 (25.5) & 4 (7.8) & 51 \\
7 & 6 (85.7) & 2 (28.6) & 1 (14.3) & 2 (28.6) & 2 (28.6) & 1 (14.3) & 1 (14.3) & 7 \\
8 & 44 (84.6) & 19 (36.5) & 3 (5.8) & 15 (28.8) & 26 (50) & 16 (30.8) & 3 (5.8) & 52 \\ \hline
3EG J1249$-$8330 & 119 (80.4) & 42 (28.4) & 14 (9.5) & 43 (29.1) & 77 (52) & 36 (24.3) & 8 (5.4) & 148 \\ \hline
\end{tabular}}
\end{center}
\vspace{-0.25 cm}
\end{table*}

We note  that almost all sources  are detected between 0.5  and 2 keV,
with half of  them also detected between 1 and 2  keV, while only very
few  sources are  detected at  very high  or very  low  energies.  The
number of sources detected in  each energy band is, in absolute terms,
very different  across the  8 pointings but,  taking into  account the
uneven effective exposure  times (as it is evident in  the case of the
observation  of field 7,  see Table~\ref{observations}),  the relative
number is constant, especially for those pointings associated with the
same {\em  EGRET} error  box. However,  we note that  below 1  keV the
fraction of detected sources is  indeed lower for the 3EG J1249$-$8330
error box than  for the 3EG J0616$-$3310 one,  probably because of the
difference   in   the   neutral   hydrogen   column   density, $\sim 10^{21}$ cm$^{-2}$ and $\sim
2.5\cdot10^{20}$ cm$^{-2}$, respectively.
\vspace{-0.25 cm}

\subsection{Count Rate and S/N Distributions}

\begin{figure*}
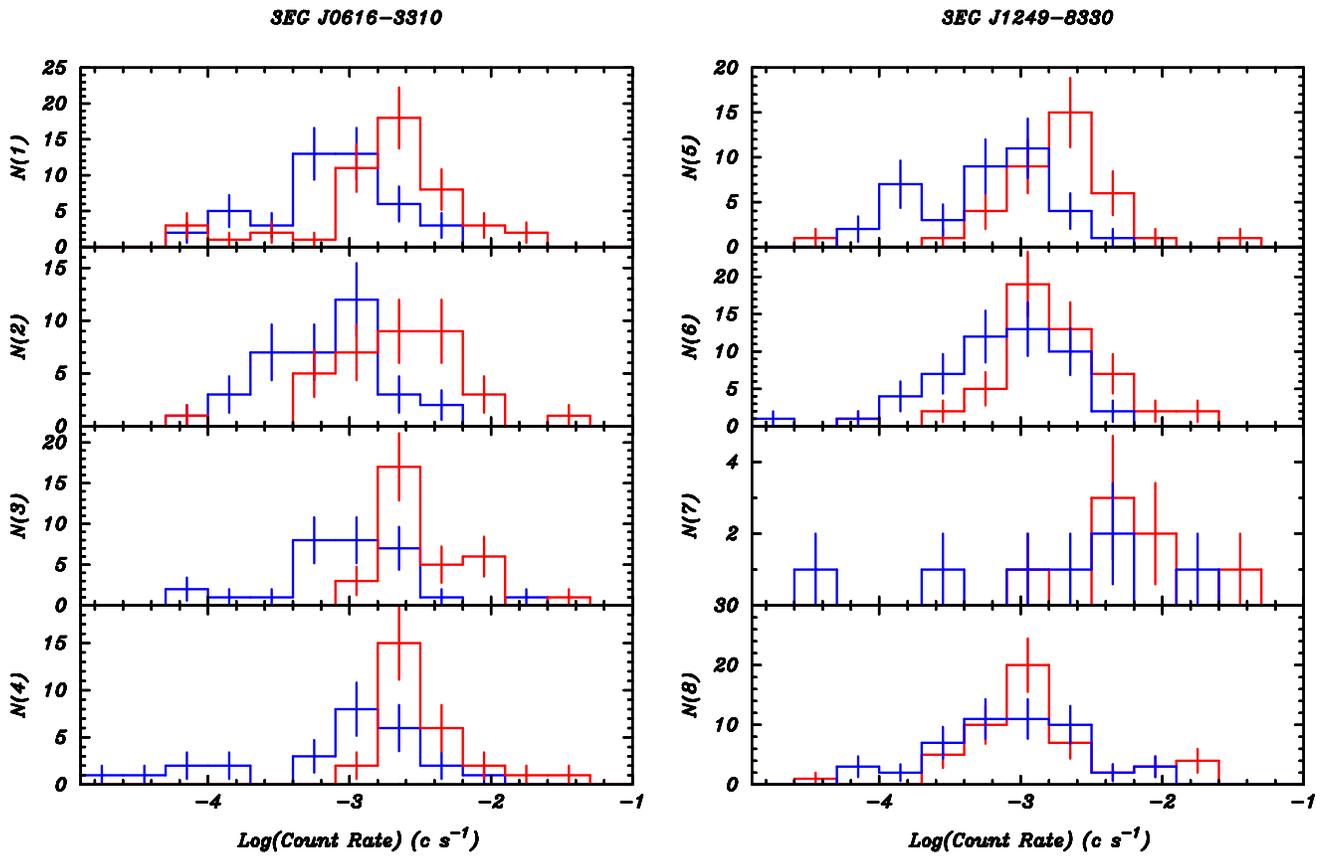

\begin{center}
\begin{tabular}{c@{\hspace{1pc}}c}
\includegraphics[height=8.5cm,angle=-90]{5247f3a.ps} &
\includegraphics[height=8.5cm,angle=-90]{5247f3b.ps} \\
\end{tabular}
\end{center}
\caption{Histograms  of the source  count--rate distributions  for each
{\em  EPIC} pointing of  the error box of 3EG  J0616$-$3310 (\textit{left, N1--N4})
and 3EG  J1249$-$8330 (\textit{right, N5--N8}),  in the energy  ranges 0.5--2  keV (\textit{red
line}) and 2--10  keV (\textit{blue line}).  For observation 7 the distribution
peaks at higher count rate  (logCR = -2.35) due to the shorter effective
exposure time.}\label{countrate}
\vspace{-0.25 cm}
\end{figure*}

The  histograms of  the source  \textit{count--rate} (\textit{CR}) distribution  for the  two
coarse soft (0.5--2  keV) and hard (2--10 keV)  energy bands are shown
in Figure~\ref{countrate},  for the single  pointings of the  two {\em
EGRET} error boxes.  Since the  number of sources per \textit{CR} bin decreases
below the  peak \textit{CR}, we deduce  that our sample  becomes incomplete for
lower    \textit{CR}    values    (the    same    approach    was    used    by
\citet{Zickgraf+97}). Therefore,  we assume the \textit{CR} peak  values as our
completeness limit. In all the  pointings of the 3EG J0616$-$3310 error
box,  the source  count--rate  distributions in  the  low (0.5--2  keV)
energy range are pretty similar, with peaks at $log~CR \approx$ -2.6; on the other hand, there are some differences among the distributions of 3EG J1249$-$8330.    The  marginal  differences  in  the
count--rate distributions  between the pointings of the two {\em  EGRET} sources are an obvious effect of the larger
hydrogen  column   density.   On   the  other  hand,   the  count-rate
distributions in  the high energy range (2--10  keV) are significantly
different, with peaks at $log~CR \le$ -3.

\begin{figure*}
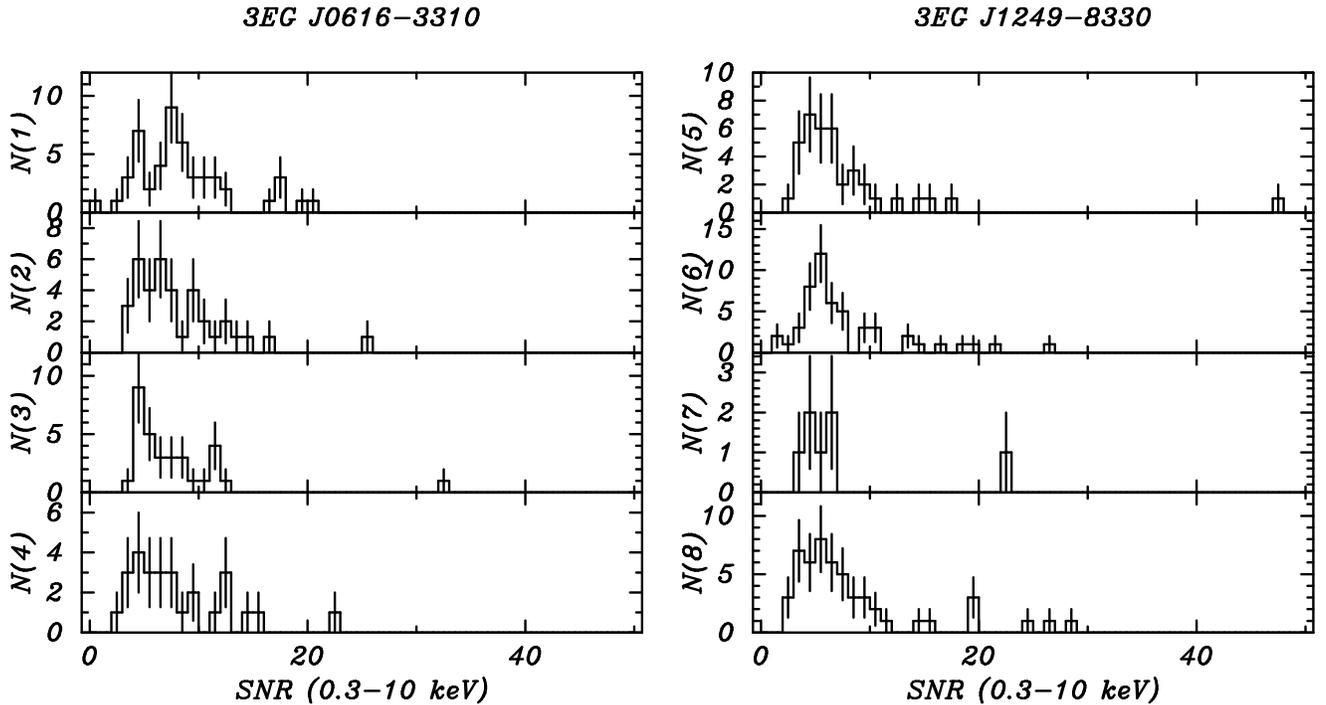

\begin{center}
\begin{tabular}{c@{\hspace{1pc}}c}
\includegraphics[height=8.5cm,angle=-90]{5247f4a.ps} &
\includegraphics[height=8.5cm,angle=-90]{5247f4b.ps} \\
\end{tabular}
\end{center}
\caption{Histogram of the source {\em S/N} distribution  for the 3EG
J0616-3310 (\textit{left, N1--N4}) and 3EG J1249$-$8330 (\textit{right, N5--N8}) fields.}\label{snr}
\vspace{-0.25 cm}
\end{figure*}

The  histograms of  the source \textit{signal--to--noise} ({\textit S/N}) distribution  in the  whole
energy  range are  reported  in Figure~\ref{snr}  for  all the
pointings of  the two {\em  EGRET} error boxes. The two  distributions peak
between 4.5 and 5.5, with only a few sources at $S/N \ge$ 10.
\vspace{-0.25 cm}

\vspace{-0.25 cm}

\subsection{Source Hardness Ratios Distribution}
\begin{figure*}[t]
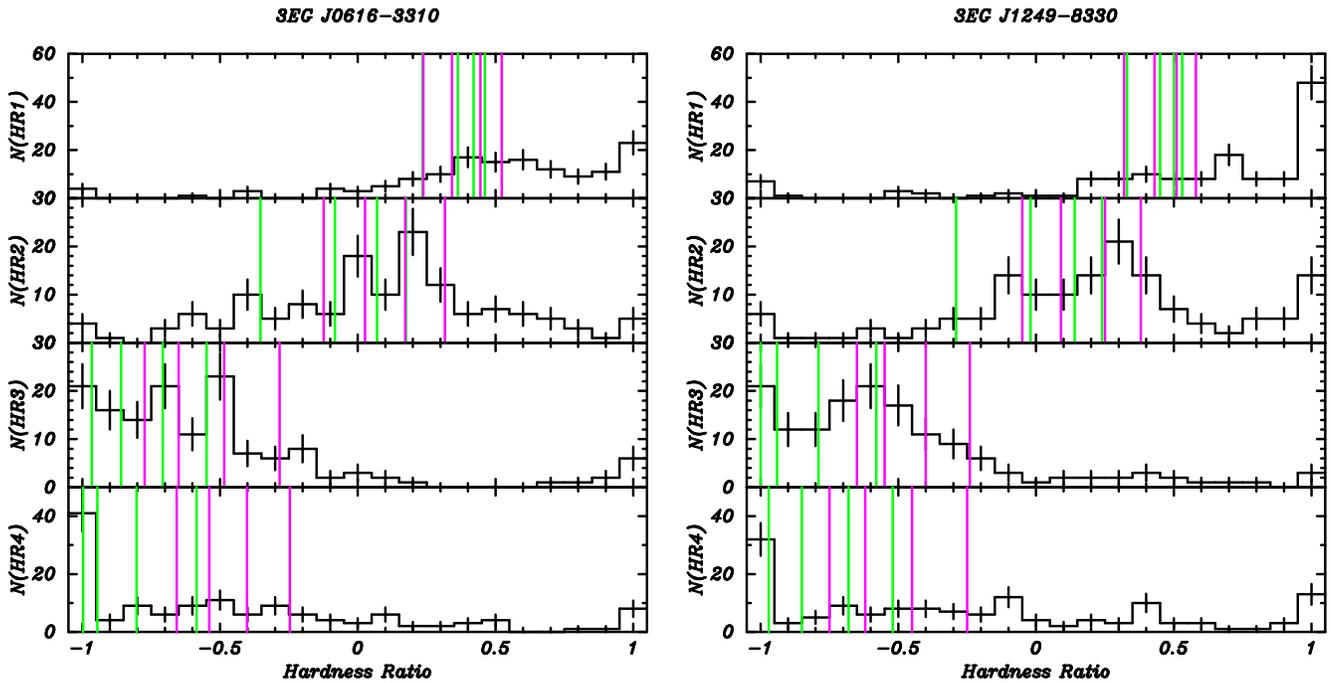

\centerline{%
\begin{tabular}{c@{\hspace{1pc}}c}
\includegraphics[width=9cm,angle=-90]{5247f5a.ps} &
\includegraphics[width=9cm,angle=-90]{5247f5b.ps} \\
\end{tabular}}
\caption{Histograms  of the \textit{HR}  distributions for  sources in  the 3EG
J0616-3310  (\textit{left}) and  3EG  J1249-8330 (\textit{right})  error boxes.  Green  bars
indicate the  expected \textit{HRs}  computed for {\em thermal bremsstrahlung}
spectra with kT = 0.5, 1, 2 and 5 keV (from left to right).  Magenta bars
indicate  the  expected  \textit{HRs}  computed  for  \textit{power--law}  spectra  with
$\Gamma$ = 1, 1.5, 2 and 2.5 (from right to left).}\label{hr}
\vspace{-0.25 cm}
\end{figure*}
\vspace{-0.25 cm}

Since  the count statistics  (usually a  few tens  of photons)  of the
detected sources  is too low  to produce significant spectra,  we have
performed  a  qualitative  spectral  analysis using  the \textit{CR}
measured  in the  seven energy  ranges defined  above to  compute four
different {\em Hardness Ratios} (\textit{HRs}):
\vspace*{0.25truecm}

\begin{small}
\noindent $HR1$ = [$CR$(0.5--1) - $CR$(0.3--0.5)]/[$CR$(0.5--1) + $CR$(0.3--0.5)]
\vspace*{0.25truecm}

\noindent $HR2$ = [$CR$(1--2)- $CR$(0.5--1)]/[$CR$(1--2) + $CR$(0.5--1)]
\vspace*{0.25truecm}

\noindent $HR3$ = [$CR$(2--4.5) - $CR$(0.5--2)]/[$CR$(2--4.5) + $CR$(0.5--2)]
\vspace*{0.25truecm}

\noindent $HR4$ = [$CR$(4.5--10) - $CR$(2--4.5)]/[$CR$(4.5--10) + $CR$(2--4.5)]
\end{small}
\vspace*{0.25truecm}

The  histograms   of  the  {\em   HRs}  distributions  are   shown  in
Figure~\ref{hr} for  the combined pointings of both  {\em EGRET} error
boxes. Most  sources   have  {\em  HR1} $\ge$ 0.5,
-0.1 $<$ {\em HR2} $<$ +0.4, {\em HR3} $<$ -0.5 and {\em
HR4} $\simeq$ -1. These  results, together  with those of  \S3.2 based on
the count distributions, suggest that,
for  both  error boxes,  the  source  population  is characterised  on
average by  rather soft spectra.

To obtain a further  indication on
the sources  spectra, we compared  the measured \textit{HRs} with  the expected
ones   computed   for   two   different   template   spectral   models
\citep{Giacconi+01,Barcons+02}, namely: a thermal {\em bremsstrahlung},
with temperatures kT increasing from 0.5 (left) to 5 keV (right), and a \textit{power--law}, with photon indexes $\Gamma$ increasing from 1 (right) to 2.5 (left).  In such a way  we could identify the spectral model
more appropriate  for a given  source and tentatively assign  its most
likely  spectral  parameters.  The  values  of  the  expected \textit{HRs}  are
overplotted as vertical  bars in Figure~\ref{hr}. As can  be seen, the
distributions are compatible with  a rather wide range of temperatures
and  photon  indexes, thus  suggesting  that  we  are indeed  sampling
different types of sources.   This conclusion is not surprising, since
the  two areas  are  at  medium galactic  latitude  and therefore  are
expected to contain both galactic and extragalactic X--ray sources.
\vspace{-0.25 cm}

\subsection{Sky Coverage and log$N$--log$S$}\label{sourceflux}

\begin{figure*}[t]
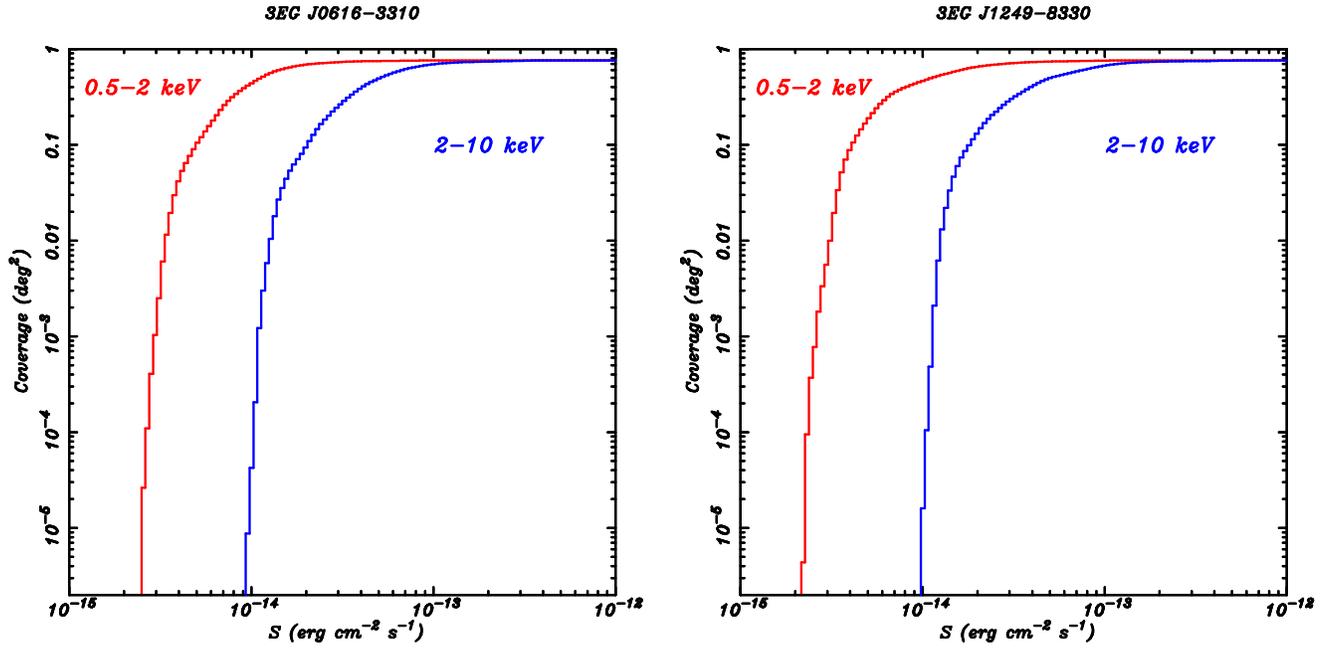

\centerline{%
\begin{tabular}{c@{\hspace{1pc}}c}
\includegraphics[height=8.5cm,angle=-90]{5247f6a.ps} &
\includegraphics[height=8.5cm,angle=-90]{5247f6b.ps} \\
\end{tabular}}
\caption{The total sky coverage of  the {\em EPIC} observations of the
3EG J0616$-$3310 (\textit{left}) and 3EG J1249$-$8330 (\textit{right}) error boxes, in the energy
bands   0.5--2    keV   (red    line)   and   2--10    keV   (blue
line).}\label{sky-coverage}
\end{figure*}

In  order to compute  the source  fluxes, we  assumed a  template {\em
power--law}  spectrum with average photon index  $\Gamma$ = 1.7. For each
pointing, we  estimated the hydrogen  column density $N_{\rm  H}$ (see
Table~\ref{observations})  from   the  relation  $N_{\rm  H}=4.8\times
10^{21}\times  E(B-V)$ cm$^{-2}$~\citep{Zombeck90}  using  the average
colour  excess   $E(B-V)$  derived   from  the  dust   maps  developed
by~\citet{Schlegel+98}.     Following    the    procedure   used    by
\citet{Baldi+02},  {\em count-rate-to-flux}  convertion  factors ({\em
CF})  were  computed  for the  {\em  PN}  and  the {\em  MOS}  cameras
individually using  their updated response matrices  and then combined
with the effective exposure times  of each pointing to derive the {\em
total} conversion  factor {\em  CF}.  For each  pointing and  for each
energy  range  we  used the  exposure  maps  of  each camera  and  the
background map  of the  merged image to  compute the {\em  flux limit}
map, which gives, at each  celestial coordinate, the minimum flux of a
source in  order to be  detected with a  probability $P=2\cdot10^{-4}$
\citep{Baldi+02}.  Then,  we used  the flux limit  maps to  derive the
total sky coverage  of both {\em EGRET} error  boxes.  These are shown
in Figure~\ref{sky-coverage},  in the two standard  {\em soft} (0.5--2
keV) and {\em hard} (2--10 keV) energy bands.

\begin{figure*}[t]
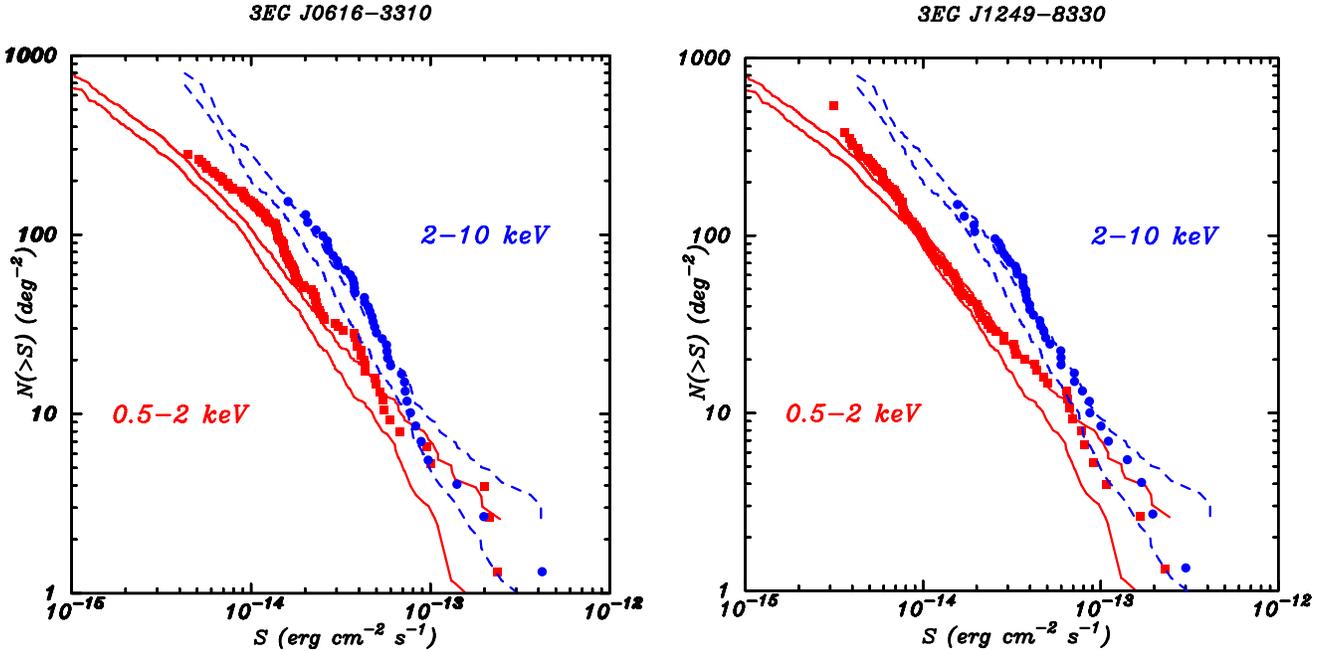

\centerline{%
\begin{tabular}{c@{\hspace{1pc}}c}
\includegraphics[height=8.5cm,angle=-90]{5247f7a.ps} &
\includegraphics[height=8.5cm,angle=-90]{5247f7b.ps} \\
\end{tabular}}
\caption{Cumulative log$N$--log$S$ distribution of the sources detected in the
{\em EPIC} observations of the 3EG J0616$-$3310 (\textit{left}) and 3EG J1249$-$8330 (\textit{right}) error boxes, in
the energy bands 0.5--2 keV (red squares) and 2--10 keV (blue dots).  For comparison, in
both diagrams we also show the lower and upper limits of the high latitude
log$N$--log$S$ in the energy ranges 0.5--2 keV (red solid lines) and 2--10 keV
(blue dashed lines).}\label{logNlogS}
\end{figure*}

Figure~\ref{logNlogS} shows
the cumulative log$N$--log$S$ distributions of the sources detected in
the  two energy  ranges  0.5--2  and 2--10  keV.   For comparison,  we
overplotted the  lower and upper  limits of the same  distributions as
measured at  high galactic latitude, where  only extragalactic sources
are  expected  to  contribute  \citep{Baldi+02}. In both cases our source density is above the upper limit of the high latitude distribution, expecially in the soft energy band and at low fluxes. This result points to a significant excess  of galactic sources, whose fraction is larger at softer energies and lower fluxes.

\vspace{-0.25 cm}

\vspace{-0.25 cm}

\section{Optical Observations}\label{sec:4}

%\subsection{ESO Observations}

In order to search for  the optical counterparts of our X--ray sources,
we       used       the       {\em       Wide       Field       Imager
(WFI)}\footnote{http://www.ls.eso.org/lasilla/sciops/2p2/E2p2M/WFI/}   mounted
at the 2.2m ESO/MPG telescope  at La Silla (Chile), where we collected
data  under programme ID  68.D--0478. The  {\em WFI}  is a  wide field
mosaic camera,  composed of eight $2048\times4096$ pixel  CCDs, with a
scale   of  $0\farcs238$/pixel   and   a  full   field   of  view   of
$33\farcm7\times 32\farcm7$.  As it  matches very well the diameter of
{\em XMM-Newton/EPIC} field of view  ($\sim$ 30 arcmin), {\em WFI} can
provide a complete coverage of the targets' area with a minimum number
of pointings.   Observations in the  filters $U (877)$, $B  (842)$, $V
(843)$,             $R            (844)$             and            $I
(845)$\footnote{http://www.ls.eso.org/lasilla/sciops/2p2/E2p2M/WFI/filters/}
were requested in order to  maximize the optical spectral coverage and
to  optimise  the objects  classification  in  the  colour space.   To
compensate  for the interchip  gaps, for  each passband  the pointings
were  split in  sequences of  five dithered  exposures with  shifts of
35$''$  and 21$''$  in RA  and DEC,  respectively.   Observations were
performed  in  Service Mode  between  December  2001  and March  2002.
Unfortunately, bad  weather conditions as well  as scheduling problems
resulted in  a highly  incomplete and inhomogeneous  dataset and  in a
data quality in some cases far from optimal.  In particular, only four
of the  eight {\em XMM--Newton} fields  were covered by  the {\em WFI},
i.e.  fields 2, 3 and 4 of the 3EG J0616$-$3310 error box and field 8 of
the  3EG J1249$-$8330 error  box (see  Table \ref{tabdatasummary}  for a
summary of the observations).
\vspace{-0.25 cm}

\begin{table*}
\begin{center}
\caption{Summary of the optical observations performed by the {\em Wide Field Imager} of the 2.2m MPG/ESO telescope at La Silla.}
\begin{tabular}{|c|c|ccccc|} \hline
Date       & Field	& Filter & Number 	& Exposure	& Average 	& Average	\\
dd.mm.yyyy & ID   	& Name   & of Frames	& Time (s)	& Seeing	& Airmass	\\ \hline
06.03.2002 & 2    	& U      & 5      	& 2500.0	& 1.20		& 1.15		\\
06.03.2002 & 2    	& B      & 5      	& 1500.0	& 1.18		& 1.28		\\
10.02.2002 & 2    	& V      & 5      	& 2000.0	& 1.12		& 1.30		\\ \hline
05.03.2002 & 3    	& U      & 5      	& 2500.0	& 1.07		& 1.14		\\
10.02.2002 & 3    	& V      & 5      	& 2000.0	& 1.29		& 1.16		\\
05.03.2002 & 3    	& R      & 5      	& 2000.0	& 0.82		& 1.05		\\
08.03.2002 & 3    	& I      & 13     	& 3250.0	& 1.01		& 1.18		\\ \hline
05.03.2002 & 4    	& U      & 5      	& 2500.0	& 1.14		& 1.30		\\
12.12.2001 & 4    	& B      & 5      	& 1500.0	& 1.09		& 1.05		\\
12.12.2001 & 4    	& V      & 5      	& 2000.0	& 1.09		& 1.01		\\
12.12.2001 & 4    	& R      & 5      	& 2000.0	& 1.00		& 1.11		\\ \hline
10.02.2002 & 8    	& B      & 1      	& 500.0		& 1.69		& 1.70		\\
11.02.2002 & 8    	& V      & 4      	& 2000.0	& 1.64		& 1.73		\\ \hline
\end{tabular}
\label{tabdatasummary}
\end{center}
\end{table*}
\vspace{-0.25 cm}

\subsection{Data Reduction}

The  data  reduction  was  performed  with a  pre-release  version  of
\textit{THELI},  a  fully  automatic  pipeline for  the  reduction  of
optical and  near-IR imaging data obtained with  single- or multi-chip
cameras.  A  detailed description of this pipeline  and the algorithms
used  can be  found  in \citet{Schirmer+03}  and in  \citet{Erben+05}.
Here we describe only the steps relevant for the the {\em WFI} multi--chip detector,
such  as astrometric and photometric calibration, image
dedithering and coaddition.  All pre-processing steps (debiasing, flat
fielding,  superflatting etc.)   are similar  to those  performed with
single chip cameras.

\begin{figure}
\resizebox{\hsize}{!}{\includegraphics[angle=-90]{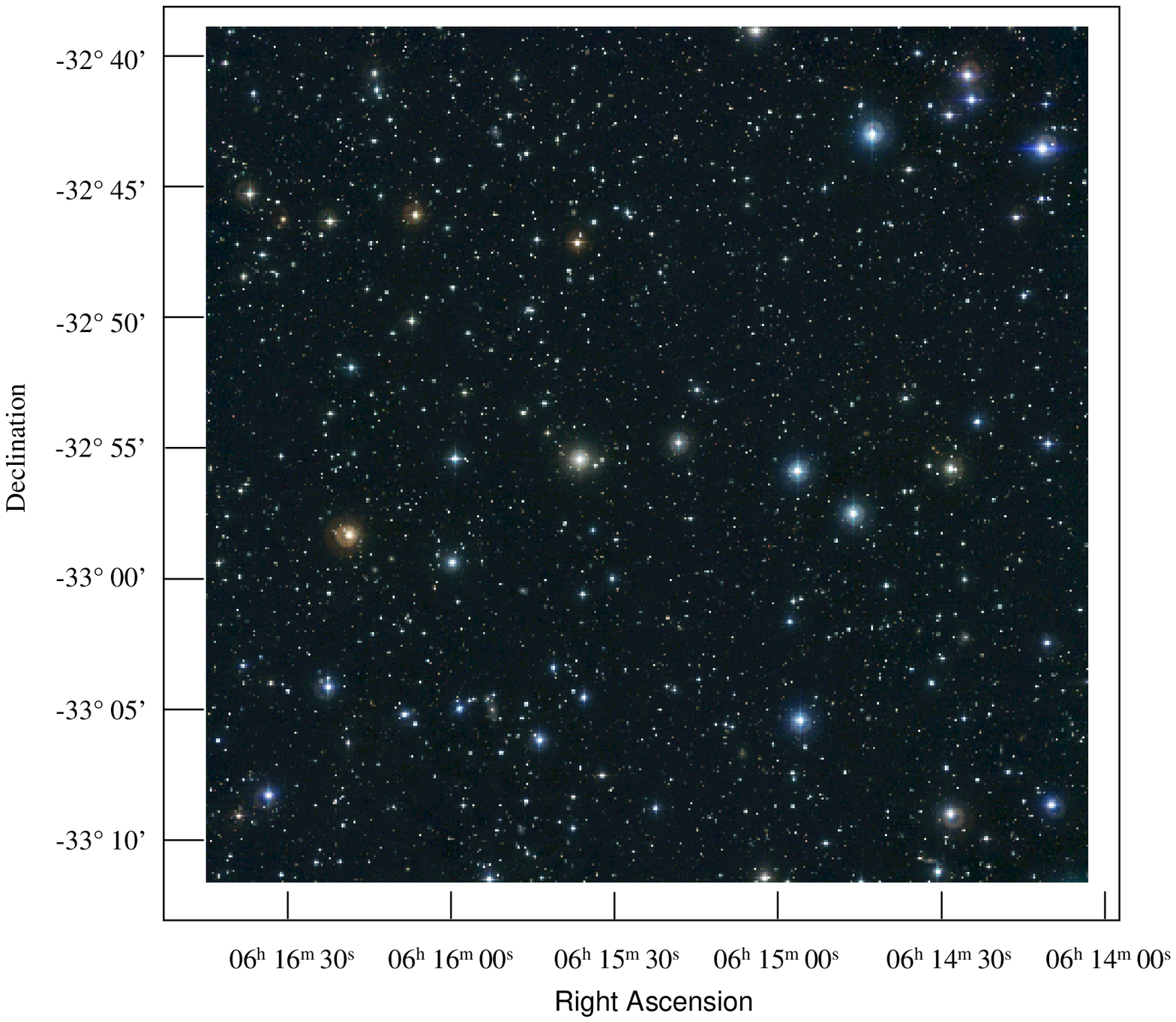}}
\caption{Composite $BVR$ image ($B$ = blue; $V$ = green; $R$ = red) of field 4 taken with {\em WFI} at the  ESO/MPG 2.2m telescope.}
\label{optical-ima}
\vspace{-0.25 cm}
\end{figure}

The astrometric  solution was performed in several  steps.  First, for
each    CCD   chip    pre--processed   image   we    run   the
\textit{SExtractor}         object         detection         algorithm
\citep{BertinArnouts96} to generate a catalogue which contained high
\textit{S/N} (DETECT\_MINAREA = 10, DETECT\_THRESH = 10) non saturated objects.
Objects  classified  as  stars  according to  the  \textit{SExtractor}
parameters  were then  matched,  using the  crude  {\em WFI}  pointing
astrometric  solution, with the  USNO--A2 astrometric  catalogue which
has  a  nominal  intrinsic  absolute  astrometric  accuracy  of  $\sim
0\farcs25$\footnote{http://ad.usno.navy.mil/star/star\_cats\_rec.html}.

Thereafter,  for  each  chip  all  the  detected  objects  within  the
overlapping regions  of the  dithering pattern were  cross--matched and
the results  were piped into  two $\chi^2$ minimisation  routines. The
first one fits two dimensional  third order polynomials to the optical
distortion  of  each  chip,  whereas  the second  one  determines  the
relative photometric  zeropoint with respect to all  other chips.  The
final image coaddition was performed chip by chip using as a reference
the  computed  chip  distorsion   maps  and  the  relative  chip  flux
normalization.

Before performing  this step, individual weight maps  were created for
each exposure, based on  normalized skyflats.  The detection of defect
pixels   was  performed   on  the   individual  chips   by   means  of
\textit{SExtractor}  and \textit{Eye}\footnote{http://terapix.iap.fr}.
Those pixels were then set to zero in the weight map.  Thereafter, the
sky  background  was  modelled  and  subtracted, and  the  chips  were
resampled    and    coadded   by    means    of   the    \textit{IRAF}
\textit{EISdrizzle} routine, using the weight maps created previously.
The registration  of the individual  chips was accurate  within $\sim$
1/10th  of a pixel  ($0\farcs02$), so  that we  did not  introduce any
artificial decrease of the  image quality.  After mosaicing, the final
image  was associated  with its  corresponding weight  map, containing
information  on  how  long  each  pixel has  been  exposed.   This  is
particular useful  since, due  to the dithering  pattern and  the gaps
between the  CCDs, the  total exposure time  in the coadded  image was
highly  inhomogeneous.  Using  the  method described  above, we  could
assign to the coadded  mosaic a consistent global astrometric solution
accurate  within  $0\farcs3$  and  a  relative  photometric  zeropoint
costant  within  $\sim$ 0.05  mag.

Figure  8  shows  as example  a
composite  {\em WFI}  image of  field 4  obtained after  co--adding the
single--passband $BVR$ reduced images.
\vspace{-0.25 cm}

\subsection{Optical Catalogues}

\begin{table*}[!ht]
\caption{Limiting   magnitudes   (3 $\sigma$)  of   our   multiband
optical--to--IR photometry. $UBVRI$ values are derived either from the
{\em  WFI} observations or, when not  available, from the
{\em GSC--2.3}  $B_{J},F,N$ equivalent (in  \textit{italics}).  $JHK$ values
are taken from {\em 2MASS}.   An hyphen indicates that no measurements
are   available.   The  last   column  lists   the  total   number  of
 objects per pointing after matching over all the passbands.}\label{multiband}
\begin{center}
\footnotesize{
\begin{tabular}{|c|cccccccc|c|} \hline
Field ID &	U   &	B          &  V   &	  R      &  I	       &   J   &  H     &K    & N \\ \hline	
1     &	-   &	{\it 23.0} &  -   &   {\it 22.0} & {\it 19.5}  &  15.8 &  15.1  &14.3 &  5688  \\
2     &	23.9&	25.1       & 24.7 &   {\it 22.0} & {\it 19.5}  &  15.8 &  15.1  &14.3 &  20837 \\
3     &	24.5&	{\it 23.0} & 24.7 &	  24.6   & 23.3        &  15.8 &  15.1  &14.3 &  32202 \\
4     &	23.8&	25.1       & 24.7 &	  24.5   & {\it 19.5}  &  15.8 &  15.1  &14.3 &  34093 \\ \hline
5     &	-   &	{\it 23.0} & -    &   {\it 22.0} & {\it 19.5}  &  15.8 &  15.1  &14.3 &  11329\\
6     &	-   &	{\it 23.0} & -    &   {\it 22.0} & {\it 19.5}  &  15.8 &  15.1  &14.3 &  7423\\
7     &	-   &	{\it 23.0} & -    &   {\it 22.0} & {\it 19.5}  &  15.8 &  15.1  &14.3 &  6820 \\
8     &	-   &	24.5       & 24.3 &   {\it 22.0} & {\it 19.5}  &  15.8 &  15.1  &14.3 &  15578\\ \hline
\end{tabular}}
\end{center}
\vspace*{-0.5truecm}
\end{table*}

Source extraction  was performed  by \textit{SExtractor} on  the final
co-added images  using the previously created weight  maps to properly
account  for  the  varying  depths  of  the  mosaics.   Our  parameter
combination  (DETECT\_THRESH = 2.0 and DETECT\_MINAREA = 8  pixels) turned
out to  be the optimal  one as it  maximizes the number  of detections
yielding  little  contamination by  spurious  objects, located  around
saturated stars.  For uniformity reasons,  we applied the  same values
for  all   catalogues.  Even  thought  the   detection  threshhold  is
admittedly  low, we  decided to  use this  value in  order  to provide
candidate optical counterparts for  as many X--ray sources as possible.
The  extracted  catalogues  were  overlayed  on  the  images  and  the
counterparts  were  visually  inspected.  The percentage  of  spurious
objects at the images' borders was less than 1\%.

Since  a very  high photometric  accuracy is  not critical  for the
purpose of  this work,  default zeropoints provided  by the  {\em WFI}
science operation team\footnote{http://www.ls.eso.org/lasilla/sciops/2p2/E2p2M/WFI/zeropoints/}
were used for the  photometric calibration. The adopted zeropoints (in
the \textit{Vega} magnitude system) are  21.96, 24.53, 24.12, 24.43  and 23.37
for the $U$, $B$, $V$, $R$, and $I$ filters, respectively.

Since the optical  coverage of the two fields  performed with the {\em
WFI} is  not complete, both  in terms of  sky and passband  (see Table
\ref{tabdatasummary}), we  have complemented our dataset  using a, yet
unpublished,  version  of  the  {\em  Guide Star  Catalogue  2}  ({\em
GSC--2.3})  which  provides  photometry  in  the  $B_J$,  $F$  and  $N$
passbands, overlapping  the Johnson's  $B$, $R$, and  $I$\footnote{see
http://www.stsci.edu/  for a  more detailed  description of  the  {\em GSC--2}
photometric system}, down to  3 $\sigma$ limiting magnitudes of $B_{J}
\sim 23$, $F \sim 22$ and  $N \sim 19.5$, with average errors of $\sim
0.25$ (at $B_j \sim 20$). In addition, to extend our passband coverage
to  the near--IR,  we have  used  the {\em  Two Micron  All Sky  Survey
(2MASS)} catalogue \citep{Cutri+03},  which provides photometry in the
$J$ ($1.25 \mu$), $H$ (1.65 $\mu$) and $K$ (2.65 $\mu$) passbands down
to limiting magnitudes of 15.8, 15.1 and 14.3, respectively.

We then cross--correlated all the available catalogues ({\em WFI}, {\em
GSC--2.3} and  {\em 2MASS})  to produce master  \textit{optical--to--IR} catalogues
for each of the eight fields.  The four {\em WFI} catalogues (relative
to fields 2,  3, 4 and 8) were cross--correlated in  turn with both the
{\em 2MASS}  Point Source and  Extended Source Catalogues  through the
{\em                             Gator}                            www
interface\footnote{http://irsa.ipac.caltech.edu/applications/Gator/}.
Then, the two output  catalogues were merged and cross--correlated with
the {\em GSC--2.3}.  For the
remaining  fields, we  extracted the  corresponding  object catalogues
from   the  {\em   GSC--2.3}   through  the   same   interface  and   we
cross--correlated  them with  both  the {\em  2MASS}  Point Source  and
Extended Source Catalogues using {\em Gator}.  In all cases, we used a
fixed cross--correlation radius of  1 arcsec which largely accounts for
the uncertainty  in the absolute astrometric calibrations  of the {\em
WFI} catalogues  ($\approx 0\farcs3$),  of the {\em  GSC--2.3} ($\approx
0\farcs35$\footnote{http://www-gsss.stsci.edu/gsc/gsc2/GSC2home.htm})
and                {\em                2MASS}                ($\approx
0\farcs25$\footnote{http://spider.ipac.caltech.edu/staff/hlm/2mass/overv/overv.html}).
The final colour  coverage for each of the  eight fields is summarized
in  Table~\ref{multiband},   together  with  the   estimated  limiting
magnitude  in each  passband and  the total  number of  single objects
extracted  from  the  optical--to--IR  master  catalogues.   The  same
catalogues  are also  used for  a  multi--colour analysis  aimed at  the
characterization  of the  stellar/galactic populations  in  the fields
(Hatziminaoglou et al., in preparation).
\vspace{-0.25 cm}

\section{Cross-correlations}\label{searchWFI}

\subsection{X--rays vs. Optical Catalogues}

\begin{table*}[!ht]
\caption{Results of the cross--correlations between the list of X--ray
sources and  the optical/IR master catalogues. For  all fields, the
total number  of candidate  counterparts is larger  than those  of the
X--ray sources because of multiple matches.}\label{statisticsWFI}
\begin{center}
\begin{tabular}{|c|c|cccc|} \hline
Field ID	& Detected	& X--ray sources	& X--ray sources	& Candidate	& Reliability	\\
		& Sources 	& with no counterpart	& with counterpart	& Counterparts	& (1-P)	\\ \hline
1	 	& 50		& 27 		& 23 	& 26		& 84 \%		\\
2		& 37		& 7 		& 30 	& 46		& 76 \%		\\
3		& 32		& 7 		& 25 	& 41		& 82 \%		\\
4		& 27		& 3 		& 24 	& 40		& 70 \%		\\ \hline
Total           & 146		& 44		& 102 	& 153		& -		\\ \hline
5		& 38		& 18 		& 20 	& 21		& 71 \%		\\
6		& 51		& 34 		& 17 	& 19		& 80 \%		\\
7	        & 7		& 5		& 2 	& 2		& 81 \%		\\
8		& 52		& 24 		& 28 	& 37		& 76 \%		\\ \hline
Total           & 148		& 81 		& 67 	& 79		& -		\\ \hline
\end{tabular}
\end{center}
\end{table*}
\vspace{-0.25 cm}

Since the absolute coordinate accuracy
of {\em XMM--Newton} is  $\sim$ 5$''$ \citep{Kirsch+04}, i.e.  a factor
10  worse  than  the  astrometric  accuracy ($\le  0\farcs5$)  of  our
optical/IR  data, we first  tried to
improve  the accuracy  of  the  X--ray   coordinates.
After overlaying  the  X--ray  positions  on the  {\em
Digital Sky Survey}  images, we found few  X--ray sources which could
be confidently associated  with a single bright optical  object and we
assumed the optical positions as  the true ones.  Then, using the {\em
IRAF} task {\em geomap}, we calculated the correction to be applied to
the X--ray coordinates (value always $\le 2''$) and we applied it to
all  the  remaining X--ray  sources  using  the  {\em IRAF}  task  {\em
geoxytran}. In the  following, we decided to use  a conservative value
of 5$''$  (i.e.  $\sim  3$ times our  astrometric correction)  for the
cross--correlation radius.

In  Table~\ref{statisticsWFI}  we report,  for  each {\em  XMM--Newton}
field, the number of X--ray sources with and without candidate
optical/IR  counterparts, as  well as  the total  number  of candidate
counterparts.    It    is   evident   that   the    results   of   the
cross-correlations  are  significantly  different  for  the  different
fields,  depending on the  varying limiting  magnitude of  the optical
coverage  (see Table~\ref{multiband}).   For instance,  for  most X--ray
sources of fields 1 and  5--7 we found no candidate counterpart within
5$''$,  owing to  the limiting flux of {\em GSC--2.3} which is, on  average, a
factor 6 shallower in flux than the {\em WFI} catalogues.  Indeed, the
fraction  of X--ray  sources without  candidate  counterparts decreases
drastically for fields  2--4.  This is particularly true  for field 4,
probably due  to its  short X--ray effective  exposure time  (see Table
\ref{observations}) which results in  the detection of only relatively
bright X--ray  sources, with presumably  brighter optical counterparts.
Conversely,  the  longer  X--ray  effective exposure  time  of field 8  results in  the detection  of fainter
X--ray   sources,   presumably   characterized   by   fainter   optical
counterparts.

The  cross--correlation   between  X--ray  and   optical  catalogues  is
obviously affected by  spurious matches.  In order to  estimate it, we
computed  from the optical/IR  master catalogues  the total  number of
objects within the areas  encompassed by each {\em XMM--Newton} fields.
Then, we used the relation  $P=1-e^{-\pi r^{2} \mu}$, where $r$ is the
X--ray error circle radius (5$''$) and $\mu$ is the
surface density of the optical objects (per square  degree), to obtain the chance coincidence
probability    between    an    X--ray    and   an    optical    object
\citep{Severgnini+05}.  By  assuming an  X--ray error circle  of 5$''$,
for each observation we estimated  the probability $P$ to vary between
16 and 30 \% over all the eight fields (see Table~\ref{statisticsWFI}),
which means  that, at  our limiting magnitudes,  contamination effects
cannot be ignored in the evaluation of the candidate counterparts.
\vspace{-0.25 cm}

\subsection{X--rays vs. Radio catalogues}\label{searchradio}

All X--ray sources  were also cross--correlated with radio catalogues, namely:

\begin{itemize}
\item  The  {\em  Parkes  Radio Source}  catalogue  (PKSCAT90),  which
includes 8264 radio object at $\delta<+27^{\circ}$ \citep{WrightOtrupcek90}
\item The  {\em Parkes--MIT--NRAO} source  catalogue (PMN) for  both the
Southern and  the Zenith  surveys: the first  one reports  23277 radio
sources  at $-87.5^{\circ}<\delta<-37^{\circ}$  \citep{Wright+94}; the
second one includes 2400 sources with $-37^{\circ}<\delta<-29^{\circ}$
\citep{Wright+96}
\item The {\em NRAO VLA Sky Survey} catalogue (NVSS), which has almost
2 million sources at $\delta>-40^{\circ}$ \citep{Condon+98}
\end{itemize}

We obtained a positional coincidence  with a NVSS object
for just  3 X--ray sources  detected in field  1: NVSS 5228  for source
XMMU J061759.1-325850, NVSS 5190/MRC 0616-329 for  source XMMU J061756.6-324735 and NVSS 4464 for source
XMMU J061721.5-330110.  Of   these,  only  source  XMMU J061756.6-324735   has  a   candidate  optical
counterpart. Moreover,  a marginal spatial correlation  with NVSS 2588
was  obtained for  source  XMMU J061546.9-333347 of  field  3, which  has no  candidate
optical  counterpart. For  all the  other X--ray  sources  no candidate
radio counterpart was found within the 5$''$ radius error circles.
\vspace{-0.25 cm}

\section{X--ray/Optical Analysis}\label{sec:6}

\subsection{The X--ray--to--optical flux ratio Classification Scheme}

\begin{table*}[!ht]
\caption{Range of  the expected $f_{\rm X}/f_{{\rm opt}}$  flux ratios (min/max value in
logarithmic units)  for the different object classes and \textit{EPIC} fields.  The  X-ray fluxes
refer to  the full {\em  XMM}-Newton energy range (0.3-10  keV), while
the optical fluxes  refer to the Johnson's $B$  (top) and $V$ (middle)
bands and to the $B_j$ band (bottom). For each source class the slight differences in the $f_{\rm X}/f_{{\rm opt}}$ values are due to the different $N_{\rm H}$ of the various \textit{EPIC} fields, which affects the count--rate--to--flux conversion factor.}\label{Fx_opt}
\begin{center}
\begin{tabular}{|c|cccccc|} \hline
Field ID	& Stars		& WDs		& CVs		& Galaxies	& Clusters	& AGN	\\
	& min/max & min/max & min/max & min/max & min/max & min/max \\ \hline
1	& -3.90/+0.50	& +1.00/+2.70	& -1.02/+1.42	& -0.54/+1.83	& +0.38/+2.11	& +0.25/+1.72	\\
2	& -3.91/+0.49	& +0.99/+2.69	& -1.02/+1.41	& -0.54/+1.83	& +0.37/+2.10	& +0.26/+1.73	\\
3	& -3.93/+0.47	& +0.95/+2.65	& -1.04/+1.39	& -0.55/+1.82	& +0.38/+2.11	& +0.23/+1.70	\\
4	& -3.94/+0.45	& +0.89/+2.60	& -1.06/+1.38	& -0.56/+1.81	& +0.38/+2.11	& +0.22/+1.69	\\
5	& -3.80/+0.60	& +1.11/+2.81	& -0.91/+1.52	& -0.44/+1.94	& +0.48/+2.21	& +0.35/+1.82	\\
6	& -3.81/+0.59	& +1.10/+2.80	& -0.92/+1.51	& -0.45/+1.92	& +0.46/+2.19	& +0.34/+1.81	\\
7	& -3.84/+0.56	& +1.04/+2.74	& -0.96/+1.48	& -0.47/+1.91	& +0.46/+2.20	& +0.32/+1.78	\\
8	& -3.78/+0.62	& +1.12/+2.82	& -0.90/+1.54	& -0.42/+1.95	& +0.49/+2.22	& +0.37/+1.84	\\ \hline
\end{tabular}
\vspace{0.25cm}

\begin{tabular}{|c|cccccc|} \hline
Field ID	& Stars		& WDs		& CVs		& Galaxies	& Clusters	& AGN	\\
	& min/max & min/max & min/max & min/max & min/max & min/max \\ \hline
1	& -4.33/+0.87	& +0.65/+3.11	& -1.54/+1.63	& -0.89/+2.04	& +0.07/+2.00	& -0.10/+2.25	\\
2	& -4.33/+0.86	& +0.64/+3.10	& -1.55/+1.62	& -0.89/+2.04	& +0.06/+2.00	& -0.09/+2.26	\\
3	& -4.35/+0.84	& +0.60/+3.06	& -1.57/+1.60	& -0.90/+2.03	& +0.07/+2.00	& -0.12/+2.23	\\
4	& -4.37/+0.83	& +0.55/+3.01	& -1.58/+1.59	& -0.90/+2.02	& +0.07/+2.00	& -0.13/+2.22	\\
5	& -4.22/+0.97	& +0.76/+3.22	& -1.44/+1.73	& -0.78/+2.15	& +0.17/+2.10	& +0.01/+2.36	\\
6	& -4.24/+0.96	& +0.76/+3.22	& -1.45/+1.72	& -0.80/+2.13	& +0.15/+2.08	& -0.01/+2.34	\\
7	& -4.27/+0.93	& +0.69/+3.15	& -1.48/+1.69	& -0.81/+2.12	& +0.16/+2.09	& -0.03/+2.32	\\
8	& -4.21/+0.99	& +0.78/+3.23	& -1.42/+1.75	& -0.77/+2.16	& +0.18/+2.11	& +0.02/+2.37	\\ \hline
\end{tabular}
\vspace{0.25cm}

\begin{tabular}{|c|cccccc|} \hline
Field ID	& Stars		& WDs		& CVs		& Galaxies	& Clusters	& AGN	\\
	& min/max & min/max & min/max & min/max & min/max & min/max \\ \hline
1	& -4.19/+0.43	& +0.74/+2.65	& -1.33/+1.31	& -0.81/+1.72	& +0.13/+1.91	& -0.02/+1.70	\\
2	& -4.19/+0.43	& +0.73/+2.64	& -1.33/+1.30	& -0.81/+1.72	& +0.12/+1.91	& +0.00/+1.71	\\
3	& -4.21/+0.41	& +0.69/+2.59	& -1.35/+1.28	& -0.82/+1.71	& +0.13/+1.91	& -0.04/+1.68	\\
4	& -4.23/+0.39	& +0.64/+2.55	& -1.37/+1.27	& -0.82/+1.71	& +0.13/+1.91	& -0.05/+1.67	\\
5	& -4.08/+0.54	& +0.85/+2.76	& -1.22/+1.41	& -0.70/+1.83	& +0.23/+2.01	& +0.09/+1.81	\\
6	& -4.09/+0.53	& +0.85/+2.75	& -1.24/+1.40	& -0.72/+1.81	& +0.21/+2.00	& +0.07/+1.79	\\
7	& -4.13/+0.49	& +0.78/+2.68	& -1.27/+1.37	& -0.73/+1.80	& +0.21/+2.00	& +0.05/+1.77	\\
8	& -4.07/+0.55	& +0.87/+2.77	& -1.21/+1.43	& -0.69/+1.84	& +0.24/+2.03	& +0.10/+1.82	\\ \hline
\end{tabular}
\end{center}
\end{table*}

Observations  performed with  several X--ray  missions have  shown that
different classes  of X--ray  emitters have different,  rather narrow,
ranges   of   X--ray--to--optical   flux   ratios   $f_{\rm   X}/f_{opt}$
\citep{Stocke+91,Krautter+99}.   In particular, such  a ratio  is very
high (i.e. $\ge$ 1000) for INSs while  it is lower for  all the other
classes of X--ray  sources, with no, or small,  overlapping between the
respective range of  values.  Therefore, we can use  this parameter in
order to reject or retain an X--ray source as a candidate INS.  For the
$f_{\rm X}/f_{\rm opt}$  values we considered the results  of the {\em
Hamburg/ROSAT All Sky  Survey} \citep{Zickgraf+03}, which provides the
typical  range of values  for each  class of  celestial sources,  as a
`reference'  classification scheme.   We considered  the  same objects
classes, i.e.  white dwarfs (WD), cataclysmic variables (CV), galaxies
(G),  cluster  of galaxies  (CG)  and  active  galactic nuclei  (AGN).
Moreover,  we considered  stars as  a single  class (S).  In  the {\em
Hamburg/RASS} the X-ray  flux is calculated in the  {\em ROSAT} energy
range 0.1--2.4 keV, assuming a `typical' spectral shape for each class
of sources, while the optical flux is based on their $B$ magnitude.

In  our case we  calculated the  source flux  in the  {\em XMM-Newton}
energy range 0.3--10 keV, and  assumed a common emission model for all
sources  (a \textit{power--law}  with photon  index  $\Gamma$ = 1.7 and  galactic
column  density). Moreover,  as  shown in  the  previous section,  the
passband coverage  of the {\em  WFI} observations is  incomplete, with
only $V$  available in all  fields (see Table \ref{multiband})  for 90
candidate  counterparts ($\sim$  70  \% of  the  total). However,  for
fields 2 and 4  the $B$ band limiting is deeper than  the $V$ band one
probably owing to the large QE  of the detector.   Therefore, to define
our classification scheme, we  decided to compute the measured $f_{\rm
X}/f_{opt}$  ratio for  both  the  $B$ and  $V$  bands, while, for
candidate counterparts  with only {\em GSC--2.3}  photometry, the $f_{\rm
X}/f_{opt}$ ratio was computed from the $B_{J}$ magnitude.

Moreover, for  each class  of sources we  had to correct  the $f_{\rm
X}/f_{opt}$ ratio  found in the  {\em Hamburg/RASS} by  accounting for
the {\em  XMM--Newton} detection  band, the different  assumed spectral
models and the different optical filter. To this aim, we have devised a
procedure which is described in the Appendix~\ref{Appendix1}. The computed
$f_{X}/f_{opt}$ ranges, relative to the $B$, $V$
and $B_J$  passbands, for the  different classes of X-ray  sources are
listed in Table~\ref{Fx_opt}. The reported values are in rough agreement with the results obtained by \citet{Krautter+99} and by \citet{Zickgraf+03}, except for the WD class. In this case, the discrepancy is due to the correction applied to the {\em
Hamburg--RASS} flux ratio in order to obtain the corresponding {\em XMM--Newton} one.  In
fact, in the first case a very soft blackbody spectrum (T = 50000 K) was
adopted, while we consider a rather hard ($\Gamma$ = 1.7), less realistic
power--law spectrum; moreover, the {\em ROSAT} energy range (0.1--2.4 keV) is
very suitable for the WD detection, which is not the case for the considered
{\em XMM--Newton} one (0.3--10 keV).  Both the items imply that, if the count rate of a
WD is high enough to be detected by {\em XMM--Newton}, we would assign it an unrealistic
large flux.
\vspace{-0.25 cm}

\subsection{Evaluation of the X--ray--to--optical flux ratios}

\begin{figure*}
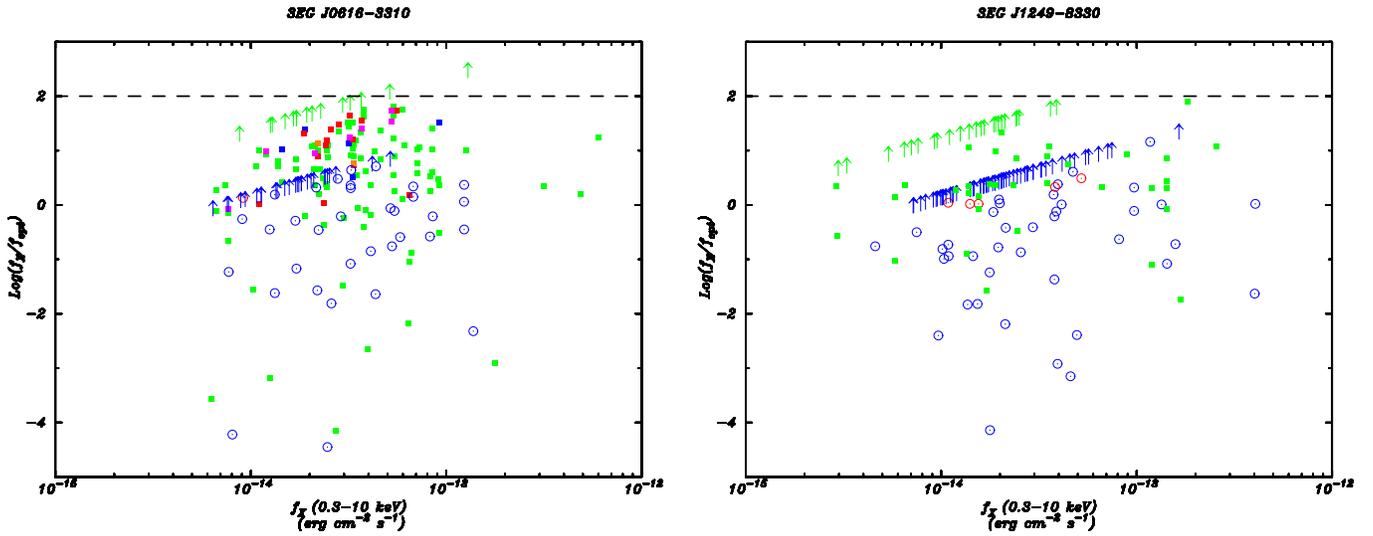

\centerline{%
\begin{tabular}{c@{\hspace{1pc}}c}
\includegraphics[width=7cm,angle=-90]{5247f9a.ps} &
\includegraphics[width=7cm,angle=-90]{5247f9b.ps} \\
\end{tabular}}
\caption{X--ray--to--optical flux ratios $f_{\rm X}/f_{\rm opt}$ for
the X--ray  sources detected  in the 3EG  J0616$-$3310 (\textit{left})  and 3EG
J1249$-$8330 (\textit{right}) fields plotted as  a function of the X--ray flux.
Different colours  refer to magnitudes computed in
different  bands   i.e.  $V$   (green),  $B,B_j$   (blue), $R,R_F$  (red).
In case of multiple optical counterparts for a single X--ray source, the $f_{\rm X}/f_{\rm opt}$ value of each candidate counterpart is shown. Vertical arrows  indicate the
lower  limit on  $f_{\rm X}/f_{\rm  opt}$ for  sources  with no  candidate
optical  counterpart  down  to  $B_{J}$ = 23, $V$ = 24.7  (\textit{left})  and
$V$ = 24.3 (\textit{right}). The dashed lines at $log(f_{\rm X}/f_{\rm opt})$ = 2 represent the threshold level for a source to be considered as a possible INS candidate.}\label{fluxratioWFI}
\end{figure*}

The values of X--ray--to--optical flux  ratios $f_{\rm X}/f_{\rm opt}$ for
all X--ray  sources are shown in Figure~\ref{fluxratioWFI}  for all the
available optical  passbands. When a single X--ray source has more than one candidate counterpart, we report the $f_{\rm X}/f_{\rm opt}$ value computed for each candidate counterpart. In  the case of  X--ray sources  with no optical counterpart,  we  estimated the lower limits  on  the  $f_{X}/f_{opt}$
ratio from  the  limiting magnitudes  of the  different
fields  (see  table   \ref{multiband}).   Thus,  we  assumed  limiting
magnitudes $V$ = 24.7 for fields 2--4 and $V$ = 24.3 for field 8 while for
fields 1 and 5--7 we assumed a limiting magnitude of $B_{J}$ = 23.

The systematic  optical identification of all X--ray  sources listed in
our  catalogue will  be  presented in  detail  in a  future paper  (La
Palombara   et  al.,   in   preparation).   However, on the basis of our
X--ray--to--optical   flux  ratio   classification   scheme  (see   Table
\ref{Fx_opt}),  it   is  very   likely  that  sources   with
$log~(f_{\rm  X}/f_{\rm opt})<$ -1.5  are stars. The
maximum  value  measured for $log~(f_{\rm  X}/f_{\rm  opt})$  is
$\sim$ 1.9 for  sources with a {\em WFI}  candidate counterpart (fields
2--4  and  8) and  $\sim$ 1.2  for those  with  a  candidate {\em  GSC}
counterpart (pontings 1 and  5--7).  These values are fully compatible
with  those  typical of  various  classes  of X--ray  sources,
especially  the  extragalactic  ones  (AGN or  cluster  of  galaxies).
Therefore, if  we assume that all  the above X--ray  sources are indeed
indentified with one of their putative optical counterparts, no matter
which one, none of them can be considered a likely INS. Of course, at this
stage  we can  not exclude  a  priori the  possibility that
some of the  candidate optical  counterparts are
just unrelated field  objects.  In this case, the X--ray sources
would  turn  out to  be  unidentified and
their corresponding  $f_{\rm X}/f_{\rm  opt}$  would increase,
moving nearer to  INS values.  Overlooked INS candidates will be
pinpointed after the systematic X--ray optical identification work, now
in progress (La Palombara et al., in preparation).
\vspace{-0.25 cm}

\section{Sources with no candidate optical counterpart}\label{sec:7}

\begin{figure*}
\includegraphics[width=18cm]{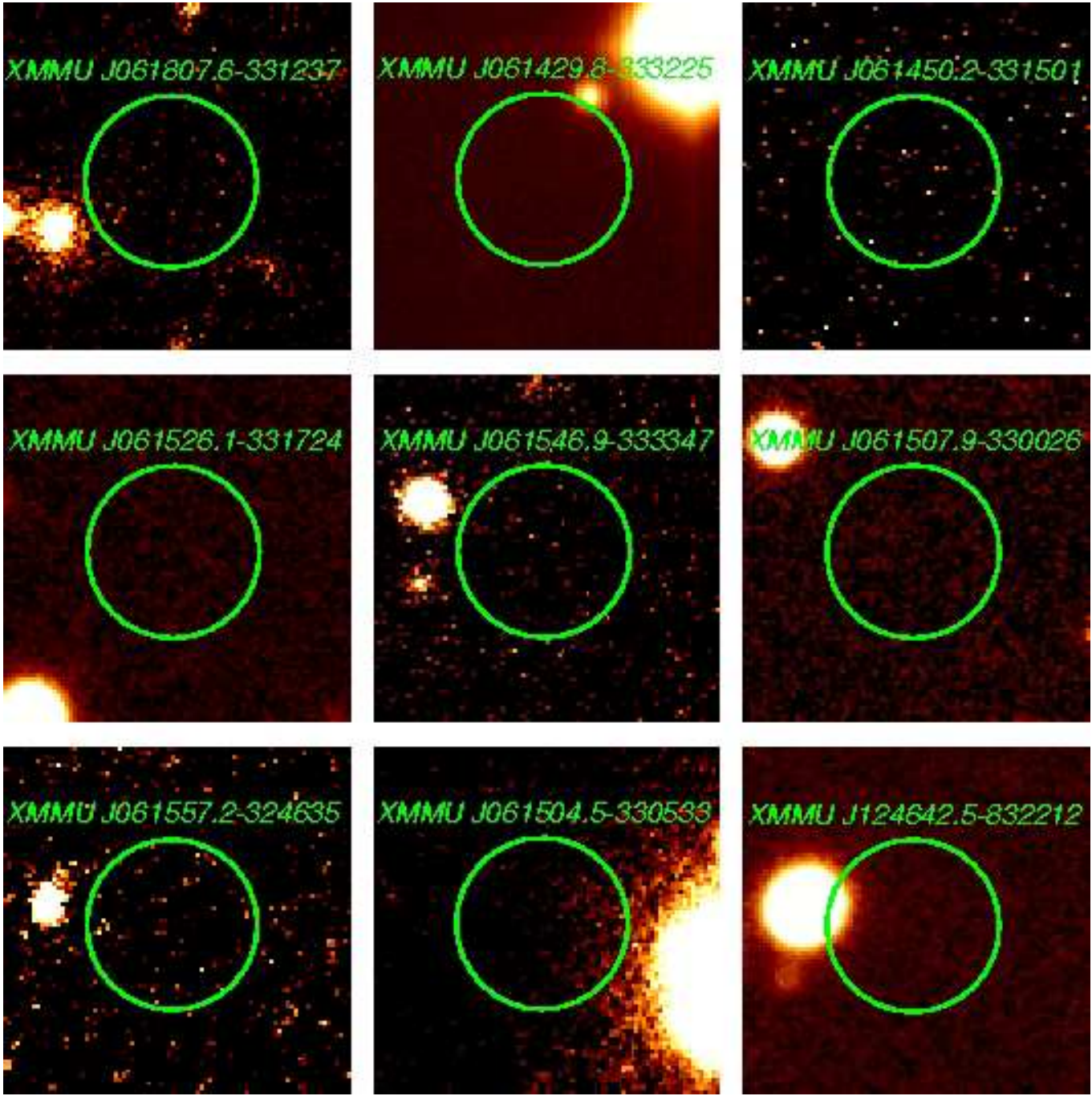}
%\centerline{%
%\begin{tabular}{ccc}
%\includegraphics[width=6cm]{chart_2_12.ps} &
%\includegraphics[width=6cm]{chart_3_32.ps} &
%\includegraphics[width=6cm]{chart_3_29.ps} \\
%\includegraphics[width=6cm]{chart_3_18.ps} &
%\includegraphics[width=6cm]{chart_3_12.ps} &
%\includegraphics[width=6cm]{chart_4_21.ps} \\
%\includegraphics[width=6cm]{chart_4_4.ps} &
%\includegraphics[width=6cm]{chart_4_22.ps} &
%\includegraphics[width=6cm]{chart_8_2.ps} \\
%\end{tabular}}
\caption{Finding charts of the X--ray sources listed in Table~\ref{candidates}. For each source the 5$''$ radius error circle is superimposed on the corresponding $V$--band \textit{WFI} image.}\label{charts}
\end{figure*}

As a first step we decided to perform a selection within our sample by
 considering  only the 125  X--ray sources  with no  candidate optical
 counterpart.   To pinpoint  more  robust INS  candidates  we have  to
 select   those  sources  which   have  the   highest  value   of  the
 X--ray--to--optical flux ratio and,  possibly, are characterized by a
 soft X--ray  spectrum, and thus  might be associated  to Geminga--like
 neutron stars. According to Table~\ref{Fx_opt}, all the typical classes of  X--ray sources are
 characterized by $log(f_{\rm X}/f_{\rm opt}) \simlt 2$, since only the brightest extragalactic sources can exceed this flux ratio level (due to the problems described in \S\ref{sec:6}, we ignore the case of the WDs). Therefore,
we have  decided to  use  $log(f_{\rm X}/f_{\rm opt})$ = 2 as a threshold value and to select the sources whose flux ratio, taking into account also the relevant uncertainties, approaches to this value. In this way we can reject all the galactic sources, almost all the galaxies and clusters of galaxies and most of the AGNs. This selection limits our sample to 9
 sources.   For  illustration  purposes, Fig.~\ref{charts}  shows  the
 positions  of  these sources  overlaied  on  the  {\it WFI}  $V$-band
 images.
%However we note that, due to the non homogeneous optical coverage, for the 125 sources 
%with no candidate  optical counterparts the  lower limits  on $log~(f_{\rm  X}/f_{\rm opt})
%$ vary between 0 and 1.4 for fields 1 and 5--7, and between 0.5 and 2.4 for fields 2--4 and 
%8. Therefore, on one hand they are not
%compelling enough to rule out non--INSs X--ray source identifications but, on the other
% hand,
% exactly because these are \textit{lower limits}, they do not rule out INS identification 
%either.
The  main   characteristics  of   these  sources  are   summarized  in
Table~\ref{candidates} where,  for each source, we  list the detection
energy bands, its flux and the corresponding X--ray--to--optical ratio
lower  limit.  We also  report  the  more  likely spectral  parameters
derived by  comparing the source \textit{HRs}  with different template
spectral  models,   namely:  a  thermal   {\em  bremsstrahlung},  with
temperatures kT$_{br}$  = 0.5, 1,  2 and 5 keV;  a \textit{blackbody},
with  temperatures  kT$_{bb}$  =  0.05,   0.1,  0.2  and  0.5  keV;  a
\textit{power--law}, with photon index $\Gamma=1-2.5$.

\begin{table*}[htbp]
\begin{center}
\caption{Main characteristics of the X-ray sources with no candidate optical counterpart and $log(f_{\rm X}/f_{\rm opt}) \simgt 2$.}\label{candidates}
\begin{tabular}{|cccccccc|} \hline
(1)	&	(2)			&	(3)				&	(4)		&	(5)		&	(6)		&	(7)		&	(8)		\\
OBS	&	SRC			&	DETECTION BANDS			&	kT$_{br}$	&	kT$_{bb}$	&	$\Gamma$	&	Flux		& 	$f_{\rm X}/f_{opt}$	\\
	&				&	(keV)				&	(keV)		&	(keV)		&			& (10$^{-14}$ cgs)	& 	(log)		\\ \hline
2	&	XMMU J061807.6-331237	&	0.5-1 				&	$\le$ 5		&	0.2		&	$\ge$1	&	2.28	$\pm$	1.23	&	1.73	$\pm$	0.31	\\ \hline
3	&	XMMU J061429.8-333225	&	0.5-2; 0.5-1; 1-2 		&	$\ge$ 2		&	0.5		&	1-2.5	&	12.92	$\pm$	3.67	&	2.48	$\pm$	0.24	\\
3	&	XMMU J061450.2-331501	&	0.5-2; 1-2 			&	2-5		&	$>$ 0.2		&	1.5-2	&	5.14	$\pm$	2.21	&	2.08	$\pm$	0.27	\\
3	&	XMMU J061526.1-331724	&	0.5-2; 1-2 			&	$\ge$ 2		&	$>$ 0.2		&	1-2	&	3.22	$\pm$	1.12	&	1.88	$\pm$	0.25	\\
3	&	XMMU J061546.9-333347	&	1-2 				&	$\ge$ 1		&	$\ge$ 0.2	&	$\le$2.5	&	1.93	$\pm$	1.03	&	1.66	$\pm$	0.31	\\ \hline
4	&	XMMU J061507.9-330026	&	0.5-2; 2-10; 1-2; 2-4.5		&	$>$ 5		&	$>$ 0.5		&	$\le$1	&	3.67	$\pm$	1.18	&	1.93	$\pm$	0.24	\\
4	&	XMMU J061557.2-324635	&	0.5-2; 1-2 			&	0.5-5		&	0.2--0.5	&	2-2.5	&	2.95	$\pm$	1.47	&	1.84	$\pm$	0.29	\\
4	&	XMMU J061504.5-330533	&	0.5-2 				&	$\le$ 5		&	$>$ 0.2	&	$\ge$1.5	&	2.06	$\pm$	1.41	&	1.68	$\pm$	0.36	\\ \hline
8	&	XMMU J124642.5-832212	&	2-10 				&	$>$ 5		&	$>$ 0.5		&	$<$1	&	3.62	$\pm$	1.69	&	1.77	$\pm$	0.20	\\ \hline
\end{tabular}
\\
\end{center}
Key to Table - Col.(1):  {\em EPIC} field sequential reference number.
Col. (2):  Source identification.   Col. (3): Source  Detection Bands,
i.e.   the energy  ranges $j$  where the  probability $P(j)$  that the
source counts  are due  to a background  fluctuation is lower  than $2
\times  10^{-4}$ (see text).   Col. (4):  Estimated range  of possible
kT$_{br}$   temperatures  for   a   thermal  bremsstrahlung   spectrum
(i.e. XSPEC {\it wabs bremss} model) compatible with the measured {\em
HRs}.  Col.  (5): Estimated  range of possible  kT$_{bb}$ temperatures
for a thermal  blackbody spectrum (i.e. XSPEC {\it  wabs bbody} model)
compatible with the measured {\em  HRs}.  Col. (6): Estimated range of
photon index $\Gamma$ values for a power law spectrum (i.e. XSPEC {\it
wabs pow} model) compatible with the measured {\em HRs}. In all cases,
we  assumed  the  galactic  $N_{\rm  H}$  estimated  in  the  pointing
direction  (Table~\ref{observations}).  The wide  range of  values for
the spectral  parameters kT$_{br}$, kT$_{bb}$ and $\Gamma$  are due to
the low  count statistics and to  the large error bars  on the derived
source {\em HRs}.  Col. (7): Estimated source flux in the energy range
0.3--10 keV, assuming a power law spectrum with $\Gamma$ = 1.7 and the
galactic $N_{\rm H}$ in the pointing direction.  Col. (8): Lower limit
(in logarithmic units) of  the X--ray--to--optical ratio, assuming the
X--ray flux of column (6) and the derived optical flux upper limits.
\end{table*}

As it is seen from Table  ~\ref{candidates}, we have singled out 8 INS
 candidate counterparts associated to  3EG J0616$-$3310.  While all of
 them  are potential  candidates, XMMU  J061807.6-331237  is certainly
 more interesting as it meets both our selection criteria, since it is
 characterized  both  by a  relatively  high X--ray--to--optical  flux
 ratio ($>  1.73 \pm  0.31$) and  by a soft  emission spectrum  (it is
 detected  only  below  1  keV).   Therefore  this  source  appears  a
 promising candidates for a Geminga-like INS.  On the other hand, XMMU
 J061429.8-333225 and XMMU  J061450.2--331501, which are both detected
 in \textit{EPIC}  field \# 3, stand  out as the only  sources with an
 X--ray--to--optical flux  ratio clearly greater  than 2 ($>  2.48 \pm
 0.24$ and  $> 2.08 \pm  0.27$, respectively).  However,  both sources
 are detected up to 2 keV while  they are not seen below 0.5 keV. As a
 result,  their  hardness   ratios  correspond  to  a  \textit{thermal
 blackbody}   temperature    greater   than   0.2   keV    or   to   a
 \textit{power--law}  photon  index $\Gamma  =  1-2.5$.  These  values
 would point towards younger objects than Geminga--like INSs.  Indeed,
 \citet{BeckerTrumper97} have shown  that the power--law components of
 X--ray  detected  INSs  have  average  photon index  of  $\sim  1.9$,
 compatible with those estimated for sources XMMU J061429.8-333225 and
 XMMU J061450.2-331501.  Thus, we consider XMMU J061807.6-331237, XMMU
 J061429.8-333225  and XMMU  J061450.2-331501 our  best  INS candidate
 counterparts  to 3EG  J0616$-$3310.  Owing  to the  shallower optical
 coverage   of   the   3EG   J1249$-$8330   error   box   (see   table
 \ref{multiband}), we could single out only one possible INS candidate
 counterpart (XMMU J124642.5-832212). The X-ray spectral parameters of
 this source are unconstrained but  they suggest a rather hard thermal
 spectrum which would not be compatible either with a Geminga--like or
 with  a young INSs.  However, the  X--ray--to--optical flux  ratio of
 this  source  ($>  1.77  \pm  0.02$),  similar to  the  one  of  XMMU
 J061807.6-331237, still makes it a possible INS candidate.
\vspace{-0.25 cm}

\section{Summary and conclusions}\label{sec:8}

Identifying  high--energy  $\gamma$--ray sources  is  a difficult  and
time--consuming task,  owing primarily to  the dimension of  the error
boxes  that have  to be  covered  at different  wavelengths.  We  have
developed  a  semiautomatic  procedure, encompassing  \textit{ad--hoc}
X--ray space observations as well  as optical ground based ones, aimed
at finding high $f_{\rm  X}/f_{\rm opt}$ candidate counterparts and we
have applied it to two middle latitude \textit{EGRET} sources. We have
mapped the error  boxes of 3EG J0616$-$3310 and  3EG J1249$-$8330 with
eight 10 ksec  {\em XMM--Newton} pointings and we  have detected about
300 X--ray  sources between  0.3 and  10 keV, down  to flux  limits of
$\sim4\times10^{-15}$ and $\sim2\times10^{-14}$ erg cm$^{-2}$ s$^{-1}$
in the energy  ranges 0.5--2 and 2--10 keV,  respectively. Four of the
eight  {\em XMM--Newton} pointings  have been  covered in  the optical
with  the {\em  Wide Field  Imager} ({\em  WFI}) at  the  2.2m MPG/ESO
telescope (La  Silla) down to a  limiting magnitude of  $V \sim$ 24.5.
For the remaining  fields, optical coverage down to  $B_J \sim 23$ has
been provided by the {\em  GSC--2.3}.  For all fields, the {\em 2MASS}
catalogue was also used to  extend the available colour coverage in the
near IR in order not  to miss possible very reddened counterparts. For
about  50  \%  of  the  detected X--ray  sources  we  found  candidate
counterparts  which yield  X--ray--to--optical flux  ratios comparable
with the typical  range of values of the  known X--ray source classes.
From the sample  of the X--ray sources with  no identification we have
selected 9 sources (8 associated to the 3EG J0616$-$3310 error box and
1 to  the 3EG  J1249$-$8330 one) characterized  by X--ray--to--optical
flux  ratios  greater than  100.   Although  all  the selected  X--ray
sources  can be considered  viable candidate  counterparts to  the two
\textit{EGRET}  sources, we  are not  yet in  the position  to propose
robust  INS  identifications.  However,  we  have  singled  out  three
interesting sources (XMMU  J061807.6-331237, XMMU J061429.8-333225 and
XMMU J061450.2-331501)  which are particularly  promising counterparts
to   3EG   J0616$-$3310    and   certainly   worth   further   optical
investigations.

While we shall pursue the study of our candidate counterparts, we note
that the need to characterize  hundreds of X--ray and optical sources,
just  to discard  them,  is an  unavoidable  bottleneck, limiting  the
efficiency  of  any multiwavelength  approach.   Since  the number  of
serendipitous,  unrelated  sources  is  proportional  to  the  surface
covered, a significant step forward will be possible only reducing the
dimension of the $\gamma$--ray  error boxes.  The next generation high
energy  gamma  ray   telescopes,  \textit{AGILE}  and  \textit{GLAST},
promise to improve the source positioning, thus significantly reducing
the uncertainty region associated to each source.  Smaller error boxes
can be covered  with less X--ray and optical  pointings, thus reducing
both  the observing  time  and the  number  of sources  in  need of  a
thorough characterization.  Hopefully, this will mark  a turning point
in the  long straggle towards the  identification of the  UGOs both as
individual sources and as a population.
\vspace{-0.25 cm}

\begin{acknowledgements}
We wish to thank M. Chieregato, A. De Luca and S. Mereghetti for their
useful comments  and suggestions. We also thank S. Ghizzardi, M. Uslenghi and S. Vercellone for their technical support. The {\em  XMM-Newton} data analysis
is supported by the Italian Space Agency (ASI), through contract ASI/INAF
I/023/05/0. The Guide Star Catalog
used  in  this  work  was  produced at  the  Space  Telescope  Science
Institute  under U.S.   Government  grant.  These  data  are based  on
photographic  data  obtained using  the  Oschin  Schmidt Telescope  on
Palomar Mountain and the UK Schmidt Telescope.  This publication makes
use of  data products from the Two  Micron All Sky Survey,  which is a
joint  project of  the University  of Massachusetts  and  the Infrared
Processing  and Analysis  Center/California  Institute of  Technology,
funded by  the National Aeronautics  and Space Administration  and the
National Science Foundation.
\end{acknowledgements}

\newpage
\bibliographystyle{aa}
\bibliography{biblio}

\appendix

\newpage
\section{Correction of the Classification Scheme}\label{Appendix1}

\begin{table}
\caption{Range of the applicable $B-V$ colours for the selected classes of objects}\label{colours}
\begin{center}
\begin{tabular}{|c|c|c|c|} \hline
CLASS	& (B-V)		& (B-V)		& REFERENCE			\\
	& min		& max		&				\\ \hline
Stars	& -0.2		& +1.8		& \citet{Hipparcos97}		\\
WDs	& -0.3		& +1.6		& \citet{Bergeron+95}		\\
CVs	& +0.2		& +2.0		& \citet{HendenHoneycutt95}	\\
Galaxies& +0.2		& +1.6		& \citet{PrugnielHeraudeau98}	\\
Clusters& +1.0		& +1.5		& \citet{Chieregato05}		\\
AGN	& -0.6		& +1.6		& \citet{Veron-CettyVeron03}	\\ \hline
\end{tabular}
\end{center}
\end{table}

According to \citet{Chieregato05}, for each class of
sources the $f_{\rm X}/f_{opt}$ in  the {\em XMM--Newton} is obtained from
the corresponding {\em Hamburg/RASS} one as follows:

\begin{equation}
\frac{f_{X,\rm XMM}}{f_{B}}=\frac{f_{X,\rm Hamburg-RASS}}{f_{B}}\times\frac{f_{X,\rm XMM}}{f_{X,\rm Hamburg-RASS}}
\end{equation}

We considered  separately the corrections due to  the different energy
ranges and to the different spectral models:

\begin{eqnarray}
\nonumber
&& \frac{f_{X,\rm XMM}}{f_{X,\rm Hamburg-RASS}}=\frac{f_{0.3-10,\rm PL}}{f_{0.1-2.4,\rm MOD}}= \\
&& \frac{f_{0.3-10,\rm PL}}{f_{0.1-2.4,\rm PL}}\times\frac{cf_{0.1-2.4,\rm PL}}{cf_{0.1-2.4,\rm MOD}}
\end{eqnarray}

For each  of the eight  {\em XMM--Newton} pointings we  estimated both
the  flux   ratio  $f_{0.3-10,\rm  PL}/f_{0.1-2.4,\rm   PL}$  and  the
conversion factor  $cf_{0.1-2.4,\rm PL}$. In addition,  for each class
of  sources   we  estimated   also  the  specific   conversion  factor
$cf_{0.1-2.4,\rm MOD}$.

The  $f_{X,\rm XMM}/f_{V}$  values were
obtained  from  the  $f_{X,\rm   XMM}/f_{B}$  ones  according  to  the
relation:

\begin{equation}
\frac{f_{X,\rm XMM}}{f_{V}}=\frac{f_{X,\rm XMM}}{f_{B}}\times\frac{f_{B}}{f_{V}}
\end{equation}

where

\begin{equation}
\frac{f_{B}}{f_{V}}=\frac{f_{B,\rm Vega}}{f_{V,\rm Vega}}\times10^{-(B-V)/2.5}=1.957\times10^{-(B-V)/2.5}
\end{equation}

For each  class of celestial source  we considered the  range of $B-V$
colours  which include  99  \% of  the  total sample (see Table~\ref{colours}).

For  candidate counterparts  with  only {\em  GSC--2.3} photometry,
we considered  the correction  factor for  the different
blue magnitude:

\begin{equation}
\frac{f_{X,\rm XMM}}{f_{B_{J}}}=\frac{f_{X,\rm XMM}}{f_{B}}\times\frac{f_{B}}{f_{B_{J}}}
\end{equation}

where:

\begin{equation}
\frac{f_{B}}{f_{B_{J}}}=\frac{f_{B,\rm Vega}}{f_{B_{J},\rm
Vega}}\times10^{-\frac{B-B_{J}}{2.5}}=0.822\times10^{-\frac{B-B_{J}}{2.5}}
\end{equation}

and $B-B_{J}=0.28\times(B-V)$  \citep{Reid+91}.  Also in  this case we
referred to Table~\ref{colours} for the $B-V$ colours of each class of
source.

\newpage
\section{Measurement of the X--ray--to--optical flux ratios}\label{Appendix2}

We computed  the $f_{X,\rm XMM}/f_{opt}$ ratios for  all the candidate
counterparts. X--ray  fluxes are  the same determined  in \S\ref{sourceflux} while
optical fluxes  were obtained from  the measured {\em  WFI} magnitudes
using  the Pogson  formula  $f_{m}=f_{m=0}\times10^{-m/2.5}$, assuming
$f_{m=0}$ = 2.904,  6.478,  3.240,  3.828  and  1.997$\times10^{6}$  erg
cm$^{-2}$  s$^{-1}$  for $m = U,  B,  V,  R$   and  $I$  passbands,
respectively \citep{Zombeck90}.

Since for  the $B_J$ band, there  is no simple relation  to obtain the
optical flux, we first estimated its Vega band flux.  This is given by
the  relation  $f_{B_{J}=0}=f_{\lambda_{\rm eff}(B_{J})}\times  W_{\rm
eff}(B_{J})$,    where   $\lambda_{\rm   eff}(B_{J})$    and   $W_{\rm
eff}(B_{J})$ are, respectively, the  effective wavelength and width of
the filter, while $f_{\lambda_{\rm  eff}(B_{J})}$ is the specific flux
calibration at $\lambda_{\rm eff}(B_{J})$ for $B_{J}=0$.  By referring
to    the    {\em   Asiago    Database    on   Photometric    Systems}
\citep{MoroMunari00} we found  $\lambda_{\rm eff}(B_{J})=4731$ \AA\ and
$W_{\rm eff}(B_{J})=1333$ \AA.  In order to obtain  the specific flux
calibration  $f_{4731}$, we  considered the  spectral  distribution of
Vega provided by \citet{Hayes85}  and performed a linear interpolation
between   4725  and   4750  \AA.    In  such   a  way   we  obtained
$f_{4731}=5.457\times10^{-9}$   erg  cm$^{-2}$   s$^{-1}$\AA$^{-1}$  and
$f_{B_{J}=0}=1333\times5.457\times10^{-9}=7.274\times10^{-6}$       erg
cm$^{-2}$   s$^{-1}$.    Finally,   we   used   the   Pogson   formula
$f_{B_{J}}=f_{B_{J}=0}\times10^{-B_{J}/2.5}$  to  obtain  the  optical
flux  for a  given magnitude  $B_{J}$.  In  the case  of the  $F$ band
magnitude, we  applied the  same procedure but  assuming $\lambda_{\rm
eff}(F)$ = 6555 \AA, $W_{\rm eff}(F)$ = 767 \AA\ and
$f_{6555} = 1.731\times10^{-9}$ erg cm$^{-2}$ s$^{-1}$ \AA\,   thus
obtaining  $f_{F}=1.328\times10^{-6}\times10^{-F/2.5}$  erg  cm$^{-2}$
s$^{-1}$.

\end{document}